\documentclass[10pt,a4paper,twocolumn]{article}

\usepackage[margin=1.8cm,columnsep=0.6cm]{geometry}
\usepackage[T1]{fontenc}
\usepackage[utf8]{inputenc}
\usepackage{lmodern}
\usepackage{microtype}
\usepackage{amsmath,amssymb,bm}
\usepackage{graphicx}
\usepackage{booktabs}
\usepackage{siunitx}
\usepackage{xcolor}
\usepackage{enumitem}
\usepackage{tikz}
\usepackage{listings}
\usetikzlibrary{arrows.meta,positioning,decorations.pathreplacing,shapes.geometric}
\usepackage[numbers,sort&compress]{natbib}
\PassOptionsToPackage{hyphens}{url}
\usepackage[colorlinks,allcolors=blue!50!black]{hyperref}
\usepackage[capitalise]{cleveref}

\graphicspath{{figures/}}

\newcommand{\code}{\texorpdfstring{\mbox{\textsf{exa-dmrg}}}{exa-dmrg}}

\definecolor{outlinecolor}{rgb}{0.10,0.25,0.50}

\newcommand{\nproc}{n_{\mathrm{p}}}
\newcommand{\cre}[1]{a^{\dagger}_{#1}}
\newcommand{\des}[1]{a^{\vphantom{\dagger}}_{#1}}

\title{\code: Toward exascale quantum chemistry with open-source DMRG, from a single CPU
core to hundreds of GPUs}
\author{\parbox{0.92\textwidth}{\centering Fabian Langkabel$^\ast$, Fabio Tarocco, Pi A.~B.~Haase,
  Fabijan Pavo\v{s}evi\'c, Stefan Knecht and
  Roberto Di Remigio Eik{\aa}s\\[6pt]
  \small Algorithmiq S.r.l., Via della Chiusa 15, 20123 Milano, Italy\\[2pt]
  $^\ast$Corresponding author: \href{mailto:fabian.langkabel@algorithmiq.fi}{fabian.langkabel@algorithmiq.fi}}}
\date{November 30, 2026}

\begin{document}

\makeatletter
\twocolumn[
  \begin{@twocolumnfalse}
    \maketitle
    \begin{abstract}
      \noindent%
The density matrix renormalization group (DMRG) is the method of choice for large active spaces in strongly correlated molecules. We present exa-dmrg, an open-source DMRG program that scales to hundreds of GPUs or thousands of CPU cores. The accuracy of DMRG is set by the bond dimension $D$, and on GPUs the largest affordable $D$ is usually limited by device memory. exa-dmrg is designed around this limit: the largest tensors are distributed across the GPUs, intermediates are formed on the fly rather than stored, and at every step the program decides which data reside on the GPUs, in host memory, or on disk. Nearly all computation is cast as batched matrix multiplications, which GPUs execute most efficiently. We demonstrate the scaling with ground-state calculations on both nitrogenase cofactors in established complete active space (CAS) models. The FeMo cofactor is treated in a (113,76) CAS with spin adaptation at $D = 10\,000$ SU(2) multiplets; on 256 GPUs a sweep takes about an hour even with tight thresholds, so a fully converged calculation completes in under a day. The P-cluster is treated in a (114,73) CAS with U(1) symmetry at $D = 28\,000$ on 512 GPUs, where exa-dmrg sustains 12.2 PFLOPS in the matrix-vector products and 9.6 PFLOPS over a full sweep, 12$\times$ and 40$\times$ the highest throughputs previously reported for DMRG (Xiang et al. 2024, https://doi.org/10.1021/acs.jctc.3c01228; Menczer et al. 2024, https://doi.org/10.1021/acs.jctc.4c00903). We also compute reactant and transition-state energies for the cytochrome P450 mediated hydroxylation of warfarin, a classic case of drug metabolism. Beyond ground states, exa-dmrg provides excited states, reduced density matrices, orbital entanglement measures, leading Slater determinants, and orbital optimization (DMRG-CASSCF), all accessible through a Python interface, directly or as an active-space solver in PySCF.
    \end{abstract}
    \vspace{1.5em}
  \end{@twocolumnfalse}
]
\makeatother

\section{Introduction}
\label{sec:intro}

The density matrix renormalization group (DMRG)~\cite{white1992,white1993} is the standard
method for strongly correlated electrons in active spaces that are too large for exact
diagonalization~\cite{white1999,chan2002,chan2011,wouters2014,olivares2015,baiardi2020}. It
represents the wave function as a matrix product state: one tensor per orbital, connected to its
neighbors in a chain by indices of dimension $D$, the bond dimension. The tensors are optimized
one after the other in sweeps along the chain. With growing $D$ the matrix product state
approaches the full configuration interaction wave function, so a single parameter controls the
accuracy, and energies can be converged or extrapolated in it.

This systematic improvability has made DMRG the method of choice for some of the hardest problems of
molecular electronic structure. Examples are iron--sulfur clusters~\cite{sharma2014} and the
cofactors of nitrogenase, the P-cluster~\cite{li2019pcluster} and the FeMo
cofactor (FeMoco)~\cite{reiher2017,li2019femoco,zhai2026}, whose many open shells pose a serious challenge to 
single-reference methods. DMRG energies also serve as the classical yardstick against which
proposals for the quantum simulation of chemistry are
measured~\cite{reiher2017,lee2023,goings2022}.

Yet, accuracy is expensive. For $K$ orbitals the cost of a sweep is dominated by $O(K^3D^3)$
operations, and the operators it works with occupy $O(K^2D^2)$
memory~\cite{white1999,chan2002}. Two demands
drive this cost up at the same time. The active space has to contain all strongly correlated
orbitals, and every further orbital in it recovers correlation that otherwise has to be added by
other means, so $K$ should be large. And the more orbitals are entangled with each other, the
larger is the $D$ that a given accuracy requires. In transition-metal clusters bond dimensions
of $10^4$ and more are needed. The most extensive study of the FeMo cofactor to date reached
$D=18\,000$ with a single sweep that took 138 hours on 24 CPU nodes, and its energy is still about
9\,kcal/mol above the extrapolated one~\cite{zhai2026}. The bond dimension that can be afforded,
in time and in memory, is what limits ab initio DMRG today.

Nearly all of the cost is dense matrix multiplication within symmetry blocks, which is the work
that graphics processing units (GPUs) do best and for which today's largest computers are built~\cite{hooker2021}.
Efficient CPU implementations parallelize DMRG on several levels at once, over the terms of the
Hamiltonian, over symmetry sectors and within the matrix multiplications, and the distribution
of the operators over the processes of a cluster~\cite{chan2004} has carried them to thousands
of cores~\cite{zhai2021,brabec2021}. GPUs entered in steps. An early implementation used a single
GPU~\cite{nemes2014}. Careful scheduling of the many small multiplications on one node with
eight GPUs sustained a quarter of a PFLOPS~\cite{menczer2025,menczer2024b,menczer2024}. The
combination of distributed operators with batched multiplications on GPUs gave the first
multi-node implementation, which reached $D=14\,000$ for the P-cluster on 48
GPUs~\cite{xiang2024}. An overview of the available DMRG programs and their capabilities is given
in a recent survey~\cite{sehlstedt2026}.

Going beyond this is mainly a question of memory. A GPU has less memory than the node that hosts
it, and the operators of a single bond can exceed both. In a distributed calculation each
process holds, besides its share of the operators, a part that does not shrink when processes
are added. The operators of every bond travel from disk through host memory to the device and
back. And the steps that remain on the host of one process, such as the Davidson iteration, do
not become faster with more GPUs~\cite{xiang2024}. Device and host memory per process then set
the largest $D$, and the parts that do not scale set the largest useful number of GPUs.

Here we present \code, an open-source program for ab initio DMRG designed for multi-node,
multi-GPU clusters. It
treats device memory as the scarce resource. The pair operators, products of two creation or
annihilation operators of a block, dominate the memory. They stay on the devices from the step
that produces them to the steps that consume them wherever they fit, and each process holds
only its own share of them. The complementary operators, which contract the operators of a
block with the two-electron integrals, are not stored at all but formed on the devices at the
bond where they are needed. The Davidson iteration and the reductions
between processes run on the devices as well. The operators of all other bonds are kept on
disk, and the host serves mainly as a staging area on the way there. At every bond the
processes agree on a memory plan, which selects for each step the fastest variant that fits
into the free device and host memory. An optional
two-dimensional distribution, over orbital pairs and over symmetry sectors, reduces the work
that otherwise stays constant per process.
\code{} supports particle-number
and spin-projection symmetry with restricted and unrestricted orbitals as well as spin
adaptation, and it also runs on CPUs alone.

We measure performance and scaling on the two cofactors of nitrogenase that have become the de
facto standard tests of large-scale DMRG, the FeMo cofactor~\cite{li2019femoco} with spin
adaptation and the P-cluster~\cite{li2019pcluster} without, and on two smaller active spaces
of cytochrome P450, which are also run on CPU nodes. On JUPITER the program
runs the P-cluster on 512 GPUs at $D=28\,000$, twice the largest bond dimension reported for
it, at a sustained rate of 9.6\,PFLOPS.

The application belongs to the same class of problems, an iron center with open shells in the
active site of an enzyme, but to a system that has been far less present in large-scale DMRG
than the nitrogenase cofactors: cytochrome P450, a family of heme enzymes that is central to
the metabolism of drugs~\cite{guengerich2008}. We study the hydroxylation of the anticoagulant warfarin by cytochrome
P450 2C9~\cite{civic2026}. The rate of such a metabolic reaction follows from its barrier, and
an error of only 1.4\,kcal/mol in the barrier changes the rate at body temperature by a factor
of ten, so the energies of the reactant and of the transition state have to be converged
tightly. DMRG calculations on models of the reactive intermediate of P450 have so far reached
43 orbitals at a bond dimension of $D = 1500$~\cite{goings2022} and 58 orbitals at a spin-adapted
bond dimension of $D = 4096$, which corresponds to about 11\,000 without spin
adaptation~\cite{menczer2024}. Here we compute reactant and transition state in an active
space of 50 orbitals built from unrestricted orbitals, at a bond dimension of $D = 7000$, and
extrapolate their energies to the limit of zero discarded weight. The same calculations in
active spaces of 76 and 100 orbitals, at bond dimensions up to $D = 12\,000$, will be added in a forthcoming revision of the manuscript once JUPITER~\cite{jupiter}, our main HPC workhorse cluster, becomes fully available again after a scheduled maintenance period.

Beyond ground-state energies, \code{} offers excited states, reduced density matrices and
orbital entanglement measures, the leading Slater determinants of a state, and orbital
optimization through an interface to PySCF~\cite{sun2017,sun2020}. The paper is organized as follows.
\Cref{sec:impl} describes the implementation, \cref{sec:bench} reports performance and scaling,
and \cref{sec:p450} presents the application to cytochrome P450. \Cref{sec:features} gives an overview of
the further capabilities, and \cref{sec:conclusion} concludes.
\section{Implementation}
\label{sec:impl}

\code{} implements the two-site DMRG algorithm in its renormalized-operator form with
complementary operators~\cite{white1999,chan2002,chan2004}. The equations of the method have been
published many times, and \cref{sec:impl-algorithm} recalls them only as far as the rest of
this section needs them. The online documentation (see Code availability) introduces DMRG from the basics
and shows for every step where and how the program implements it. The other subsections describe
what is specific to this implementation: which operators are kept and where they live, how the
two expensive steps run on the GPUs, and how operators and work are distributed over many GPUs.
Throughout, $K$ is the number of active orbitals, $D$ the bond dimension and $\nproc$ the number
of MPI ranks, each of which drives one GPU. An active space of system X with $N$ electrons in
$K$ orbitals is written X($N$,$K$). Device
memory is what limits the bond dimension in practice, so the program is built to need as little
of it as possible for given $K$ and $D$.

The description below follows the GPU path. \code{} can also run entirely on CPUs. Every step has a host
implementation that executes the same plans with threaded BLAS and LAPACK routines, on one rank
or on many, and the discussion of device memory below then applies to host memory instead.

\subsection{Algorithm in brief}
\label{sec:impl-algorithm}

\begin{figure}[!htb]
  \centering
  \resizebox{\columnwidth}{!}{\begin{tikzpicture}[font=\footnotesize, >={Stealth[length=1.6mm]}]
  \definecolor{cL}{rgb}{0.62,0.75,0.93}
  \definecolor{cR}{rgb}{0.98,0.80,0.55}
  \node[font=\footnotesize\bfseries] at (1.9,1.3) {orbital chain};
  \node[font=\footnotesize\bfseries] at (6.05,1.3) {device memory};
  \node[font=\footnotesize\bfseries] at (9.6,1.3) {disk};
  \node[cylinder, shape border rotate=90, draw=black!60, fill=black!8, aspect=0.18,
        minimum width=0.85cm, minimum height=4.5cm] at (9.6,-1.75) {};

  \foreach \r/\lab in {0/{$b$},1/{$b{+}1$},2/{$b{+}2$}} {
    \pgfmathsetmacro{\y}{-1.7*\r}
    \pgfmathtruncatemacro{\nl}{2+\r}
    \pgfmathtruncatemacro{\da}{3+\r}
    \pgfmathtruncatemacro{\db}{4+\r}
    \pgfmathtruncatemacro{\rf}{5+\r}
    \pgfmathtruncatemacro{\s}{mod(\r,2)}
    \pgfmathsetmacro{\xsrc}{4.75+1.25*\s}
    \pgfmathsetmacro{\xtgt}{4.75+1.25*(1-\s)}

    \node[anchor=west, inner sep=0pt] at (0.05,\y+0.5) {bond \lab};
    \fill[cL, rounded corners=2pt] (0.05,\y-0.2) rectangle (0.35+0.34*\nl,\y+0.2);
    \fill[cR, rounded corners=2pt] (0.05+0.34*\rf,\y-0.2) rectangle (3.75,\y+0.2);
    \foreach \i in {0,...,10} {
      \pgfmathtruncatemacro{\isdot}{(\i==\da || \i==\db) ? 1 : 0}
      \ifnum\isdot=1
        \draw[thick, fill=white] (0.2+0.34*\i,\y) circle (0.085);
      \else
        \fill[black!60] (0.2+0.34*\i,\y) circle (0.075);
      \fi
    }

    \draw[black!50, fill=black!4, rounded corners=3pt] (4.05,\y-0.42) rectangle (8.05,\y+0.42);
    \fill[cL] (\xsrc-0.55,\y-0.28) rectangle (\xsrc+0.55,\y+0.28);
    \draw[black!70] (\xsrc-0.55,\y-0.28) rectangle (\xsrc+0.55,\y+0.28);
    \node at (\xsrc,\y) {$A, B$};
    \fill[cL!35] (\xtgt-0.55,\y-0.28) rectangle (\xtgt+0.55,\y+0.28);
    \draw[black!70, dashed] (\xtgt-0.55,\y-0.28) rectangle (\xtgt+0.55,\y+0.28);
    \node[black!65] at (\xtgt,\y) {$A, B$};
    \fill[cR] (6.8,\y-0.28) rectangle (7.9,\y+0.28);
    \draw[black!70] (6.8,\y-0.28) rectangle (7.9,\y+0.28);
    \node at (7.35,\y) {$A, B$};

    \ifnum\s=0
      \draw[->, semithick] (\xsrc+0.1,\y+0.28) to[bend left=60] (\xtgt-0.1,\y+0.28);
    \else
      \draw[->, semithick] (\xsrc-0.1,\y+0.28) to[bend right=60] (\xtgt+0.1,\y+0.28);
    \fi
    \draw[->, dashed] (\xtgt+0.3,\y-0.28) |- (9.15,\y-0.62);
    \draw[->] (9.15,\y+0.05) -- (7.9,\y+0.05);
  }
  \draw[->, semithick] (5.75,-0.28) -- (5.75,-1.42);
  \draw[->, semithick] (4.50,-1.98) -- (4.50,-3.12);

  \node at (0.71,-0.42) {$L$};
  \node at (2.75,-0.42) {$R$};
  \node at (5.375,0.83) {renormalization};
  \node[anchor=east] at (5.67,-1.0) {stays on the device};
  \node[anchor=south] at (8.6,0.05) {read};
  \node[anchor=north] at (8.6,-0.64) {write};
\end{tikzpicture}}
  \caption{Operators in device memory along a forward sweep. At each bond the pair operators of
    the left block (blue) are read from the buffer that the previous renormalization wrote. The
    renormalization writes those of the grown block into the other buffer (dashed), which is
    the source at the next bond. As long as both buffers fit into device memory, these
    operators do not leave the device. A copy is written to disk in the background, and the
    operators of the right block (orange) are read from disk when their bond is prepared.}
  \label{fig:buffers}
\end{figure}
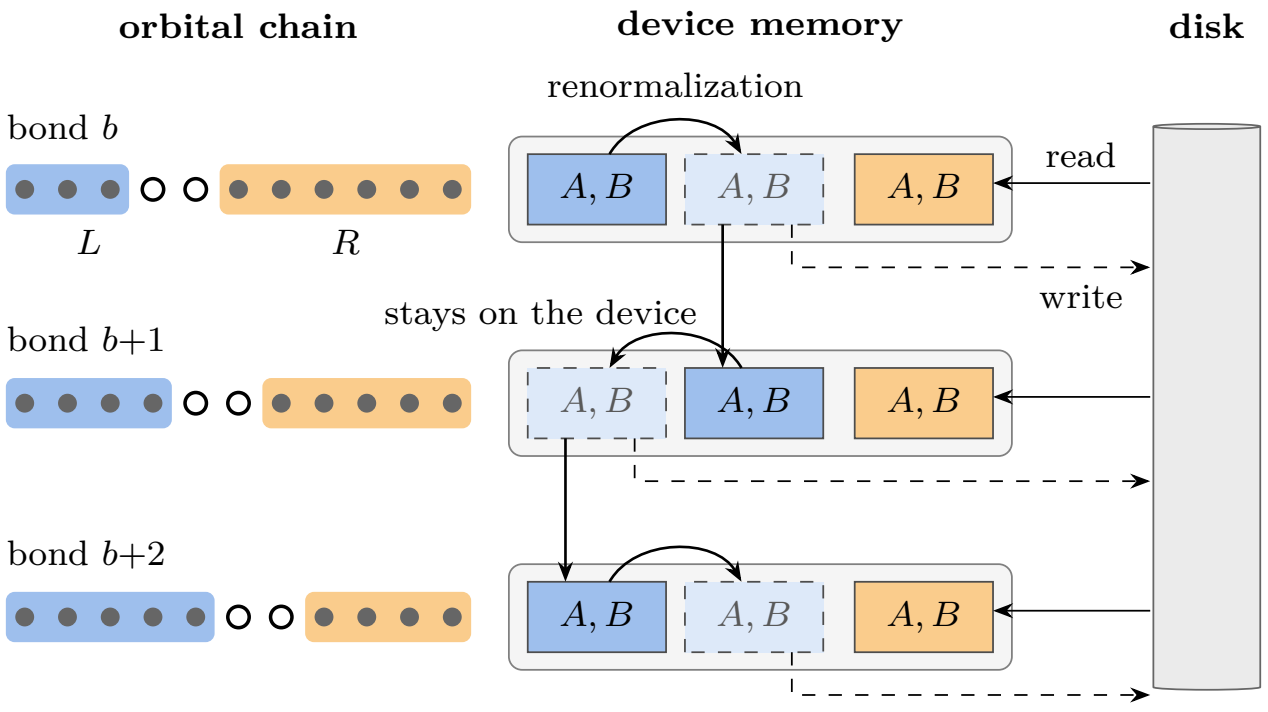

DMRG optimizes the wave function locally. The orbitals are arranged in a chain, and at each step
the chain is divided at a bond into a left block $L$, two explicit sites and a right block $R$
(\cref{fig:buffers}, left). A site is one orbital with its four states. A block stands for all
orbitals on one side of the bond. Their full space grows as $4^k$ with the number $k$ of
orbitals, so a block keeps only a basis of at most $D$ many-electron states that earlier steps
have selected. This renormalized basis is never written down. A block is known only through the
matrices of a set of operators in this basis, and these matrices are what the program stores and
transforms.

A step at a bond has four phases:
\begin{enumerate}[leftmargin=1.6em,itemsep=1pt,topsep=3pt,parsep=0pt]
  \item The complementary operators of the bond, which are defined below, are prepared.
  \item The lowest eigenvector $\Psi$ of the Hamiltonian in the product space of the four parts
    is computed with the Davidson method~\cite{davidson1975}, which applies the Hamiltonian to a
    vector, $\sigma = H\Psi$, once per iteration.
  \item The reduced density matrix of the block that grows by one site is diagonalized, and
    its $D$ eigenvectors of largest weight form the isometry $C$ onto the basis of the grown
    block (decimation).
  \item The operators of the grown block are transformed into this basis (renormalization),
    and $\Psi$ is transformed into the starting vector of the next bond~\cite{white1996}.
\end{enumerate}
A sweep visits all bonds
in one direction, and the direction alternates from sweep to sweep until the energy is converged.

Because the blocks are known only through operator matrices, the electronic Hamiltonian,
\begin{equation}
  H = \sum_{pq,\sigma} h_{pq}\, \cre{p\sigma}\des{q\sigma}
    + \frac12 \sum_{pqrs,\sigma\tau} (pq|rs)\,
      \cre{p\sigma}\cre{r\tau}\des{s\tau}\des{q\sigma},
\end{equation}
has to be written as a sum of products of operators that each act on one part. Taken term by
term, the two-electron part would give $O(K^4)$ such products. Complementary
operators~\cite{xiang1996,white1999,chan2002} reduce this number. Wherever two or three of the
four operators of a term act on the same block, they are summed, with the integrals as weights,
into a single operator of that block. With each site assigned to its neighboring block,
this gives
\begin{align}
  H ={}& H^{L} + H^{R}
     + \sum_{i\in L,\sigma}\bigl(\cre{i\sigma} S^{R}_{i\sigma} + \text{h.c.}\bigr) \notag\\
   & + \sum_{i\in R,\sigma}\bigl(\cre{i\sigma} S^{L}_{i\sigma} + \text{h.c.}\bigr) \notag\\
   & + \sum_{ij\in N}\sum_{\sigma\tau}\Bigl[\bigl(A^{\sigma\tau}_{ij}P^{\sigma\tau}_{ij}
     + \text{h.c.}\bigr) + B^{\sigma\tau}_{ij}Q^{\sigma\tau}_{ij}\Bigr].
  \label{eq:decomposition}
\end{align}
$H^{X}$ collects all terms inside block $X$. The terms with $S$ contain one operator on one
block and three on the other. In the last sum two operators act on each block. One of the
blocks, $N$, carries the normal operators $A$ and $B$, which are plain products of two
elementary operators, and the other block carries the complementary operators $P$ and $Q$:
\begin{align}
  A^{\sigma\tau}_{ij} &= \cre{i\sigma}\cre{j\tau}, \qquad
  B^{\sigma\tau}_{ij} = \cre{i\sigma}\des{j\tau}, \notag\\
  P^{\sigma\tau}_{ij} &= \frac12 \sum_{kl} (ik|jl)\, \des{l\tau}\des{k\sigma}, \notag\\
  Q^{\sigma\tau}_{ij} &= \delta_{\sigma\tau} \sum_{kl,\lambda} (ij|kl)\,
      \cre{k\lambda}\des{l\lambda} - \sum_{kl} (il|kj)\, \cre{k\tau}\des{l\sigma}, \notag\\
  S_{i\sigma} &= \frac12 \sum_{j} h_{ij}\, \des{j\sigma}
      + \sum_{jkl,\tau} (ij|kl)\, \cre{k\tau}\des{l\tau}\des{j\sigma}.
  \label{eq:operators}
\end{align}
The sums over $j$, $k$ and $l$ run over the orbitals of the block the operator acts on, and the
signs from the fermionic order of the blocks are absorbed into the operators. $N$ is chosen as
the block with fewer orbitals~\cite{chan2004,zhai2021}. \Cref{eq:decomposition} then has
$O(K^2)$ terms, and applying $H$ costs $O(K^2D^3)$ per iteration, or $O(K^3D^3)$ over a sweep.
For unrestricted orbitals the integrals carry spin labels, $h^{\sigma}_{pq}$ and
$(pq|rs)^{\sigma\tau}$, and nothing else changes.

The pair operators $A$ and $B$ make up almost all of the data of a calculation, and the way
\code{} handles them is subject to the rest of this section (\cref{fig:buffers}). The
pair operators of the growing block are produced on the device by the renormalization at one
bond. At the next bond they are consumed on the device: by the forming of $P$ and $Q$, by the
Davidson iteration and by the next renormalization. The disk keeps the operators of the bonds
that are not being solved, and the host serves mainly as a staging area between the two.
\Cref{sec:impl-operators}
describes the stored operators, \cref{sec:impl-matvec,sec:impl-renorm} the two expensive steps,
\cref{sec:impl-memory} the use of the three kinds of memory and \cref{sec:impl-mpi} the
distribution over GPUs.

\subsection{Symmetry blocks and stored operators}
\label{sec:impl-operators}

The Hamiltonian conserves the number of electrons $N$, the spin projection $S_z$ and the total
spin $S$. \code{} always exploits $N$ and $S_z$. We refer to this pair of Abelian symmetries as U(1) symmetry. The
states of every basis are sorted into sectors of equal $(N, S_z)$.
An operator changes these quantum numbers by a fixed amount. The creator $\cre{i\uparrow}$, for
example, maps the sector $(N, S_z)$ to $(N{+}1, S_z{+}\tfrac12)$ and to no other. Its matrix
therefore consists of one dense block per sector and of zeros elsewhere, and only these blocks are
stored. The same holds for the isometries and for $\Psi$, which has blocks only where the
quantum numbers of its four parts add up to those of the target state. All numerical work is
done on these dense blocks, whose dimensions are fractions of $D$. Which blocks exist follows
from the quantum numbers alone and never from the size of matrix elements, so the sequence of
operations in a run does not depend on the data. Spin-adapted calculations use multiplets of
the total spin instead (\cref{sec:impl-su2}).

\Cref{tab:operators} lists the operators of a block. Only $\cre{i\sigma}$, the pair operators
$A$ and $B$, $S$ and $H^{X}$ are kept from bond to bond. Because
$A^{\tau\sigma}_{ji} = -A^{\sigma\tau}_{ij}$ and
$B^{\tau\sigma}_{ji} = (B^{\sigma\tau}_{ij})^{\dagger}$, only the half with
$(i\sigma) \le (j\tau)$ is stored, which amounts to $4K_X^2$ operators for a block $X$ with $K_X$
orbitals.

In the standard algorithm $P$ and $Q$ are carried from bond to bond like all other
operators~\cite{white1999,chan2002}. \code{} never renormalizes or stores them. By
\cref{eq:operators} they are linear combinations of the $A$ and $B$ of their own block, with
integrals as coefficients, and the renormalization is linear as well. Renormalizing $P$ and $Q$
is therefore the same as forming them from the renormalized $A$ and $B$, and this can wait until
the bond that needs them. There,
\begin{equation}
  P^{\sigma\tau}_{ij} = \sum_{kl} W_{(ij),(kl)}\, \bigl(A^{\sigma\tau}_{kl}\bigr)^{\dagger},
  \qquad W_{(ij),(kl)} = \tfrac12\,(ik|jl),
  \label{eq:forming}
\end{equation}
is evaluated for all pairs $(ij)$ at once. For one pair of sectors, the blocks of all
$A_{kl}^{\dagger}$ are written as the rows of one matrix, and a single multiplication with the
matrix $W$ of integrals yields the blocks of all $P_{ij}$. $Q$ follows from $B$ in the same way.

This choice has two advantages. First, every block carries the same kinds of operators
wherever it lies in the chain. Second, $P$ and $Q$ exist only at the bond that is being solved and only in
device memory, so they are never written to disk or read back. There are also two costs.
First, a block keeps the $A$ and $B$ of all its orbital pairs even when it has more orbitals
than the other block. Second, the forming takes $O(K_N^2 K_C^2 D^2)$ operations per bond, where $K_N$ and $K_C$ are the numbers
of orbitals in the normal and in the complementary block. Over a sweep this adds up to
$O(K^5D^2)$, where the standard algorithm spends $O(K^4D^2)$ on updating its stored
complementary operators~\cite{white1999,chan2002}. The term remains small against the
$O(K^3D^3)$ of the matrix--vector products as long as $D$ is not small compared with $K^2$, which
is the case in the calculations the program is built for. In those of \cref{sec:bench} the
forming with its completion takes about a tenth of the time of a sweep or less in most runs and
up to 28\,\% at the largest number of GPUs of a series, because a part of this work does not shrink with
the number of GPUs (\cref{sec:impl-mpi,sec:bench-analysis}).

\begin{table}[!htb]
  \caption{Operators of a block $X$ with $K_X$ of the $K$ orbitals. An operator is either
    stored from bond to bond or formed at the bond that is solved. $K_N$ is the number of
    orbitals of the block that carries the normal operators at that bond and $n_{\mathrm{K}}$
    the number of Krylov vectors.}
  \label{tab:operators}
  \centering
  \footnotesize
  \vspace{3pt}
  \renewcommand{\arraystretch}{1.15}
  \begin{tabular}{@{}l@{\hspace{7pt}}l@{\hspace{7pt}}l@{\hspace{7pt}}l@{}}
    \toprule
    Operator & Number & Kept & Memory per rank \\
    \midrule
    $\cre{i\sigma}$ & $2K_X$ & stored & $O(K_X D^2/\nproc)$\textsuperscript{a} \\
    $A$, $B$ & $4K_X^2$ & stored & $O(K_X^2 D^2/\nproc)$ \\
    $S_{i\sigma}$ & $2(K-K_X)$ & stored & $O((K-K_X) D^2/\nproc)$ \\
    $H^{X}$ & 1 & stored & $O(D^2)$ \\
    $P$, $Q$ & $4K_N^2$ & formed & $O(K_N^2 D^2/\nproc)$ \\
    Krylov vectors & $n_{\mathrm{K}}$ & formed & $O(n_{\mathrm{K}} D^2/\nproc)$ \\
    \bottomrule
  \end{tabular}

  \vspace{2pt}
  \raggedright\footnotesize
  \textsuperscript{a}\,While a bond is prepared, a rank also holds a working copy of the
  $\cre{i\sigma}$ of both blocks, $O(KD^2)$, which does not shrink with $\nproc$.
\end{table}

\subsection{Matrix--vector product}
\label{sec:impl-matvec}

Applying $H$ to a vector is the innermost operation of the method. It is repeated in every
Davidson iteration at every bond and takes most of the time of a sweep. In block-sparse form it
is not one large matrix product but a very large number of small ones. Each of the $O(K^2)$
terms of \cref{eq:decomposition} multiplies every block of $\Psi$ that it connects, from the
left with a block of an operator on $L$ and from the right with a block of an operator on $R$. A GPU is
used well only if these products are issued in large batches and if their operands are already
in device memory. The matrix--vector product is therefore split into a preparation that runs
once per bond and an execution that runs once per iteration.

The starting point is the action of a single term. For fixed states $s_1$ and $s_2$ of the two
sites, $\Psi^{(s_1 s_2)}$ is a block-sparse matrix whose rows belong to the basis of $L$ and
whose columns to that of $R$. With Abelian symmetries a site operator connects single states of
its site, so it contributes only a number, and a term acts as
\begin{equation}
  \sigma^{(s_1' s_2')} \mathrel{+}= w\; O^{L}\, \Psi^{(s_1 s_2)}\, \bigl(O^{R}\bigr)^{\mathsf T},
  \label{eq:term}
\end{equation}
where $O^{L}$ and $O^{R}$ are the block operators of the term and $w$ collects the site matrix
elements and signs. The product is evaluated in two stages, each a multiplication of two
matrices,
\begin{equation}
  Y = \Psi^{(s_1 s_2)}\, \bigl(O^{R}\bigr)^{\mathsf T}, \qquad Z = O^{L}\, Y .
  \label{eq:stages}
\end{equation}
$Y$ and $Z$ are the intermediates of the term, and $wZ$ is its contribution to $\sigma$.
The same $Z$ results if the left operator is applied first, $Y' = O^{L}\,\Psi^{(s_1 s_2)}$
and $Z = Y'\,(O^{R})^{\mathsf T}$. Because the blocks are not square, the two orders differ in
the number of operations. The plan chooses the cheaper one for every first-stage product together
with the second-stage products that read its result. If several products add into one intermediate, they all keep the order already chosen for it. First-stage products whose result no term reads are dropped.

Many terms contribute to the same block of $\sigma$, and their contributions are not added one
by one. Each $Z$ is written to its own place in a buffer, and a reduction kernel then adds
them, each with its weight $w$, into $\sigma$. Keeping the weights out of the products has
three advantages. The operator blocks are used unscaled, so a block that occurs in many terms
is uploaded once. A first-stage product that several terms need is computed once. And products
of equal shape can be batched whatever their weights are. Terms with an identity on one side
skip a stage.

The preparation compiles these operations into a plan. The plan is the complete list of the
matrix products of both stages and of the entries of the reduction for one bond, with every
block addressed by its offset in a contiguous device buffer. It can be built before any number
is known, because the block structure follows from the quantum numbers, and all Davidson
iterations of the bond execute the same plan. The products of a stage are sorted by shape and
executed as batched general matrix multiplications (GEMMs) on concurrent streams~\cite{xiang2024}.

The intermediates $Y$ and $Z$ of \cref{eq:stages} have the size of a block of $\Psi$ each, and
those of all terms together would exceed the memory of a device many times over. The plan is
therefore cut into chunks. A chunk is a set of terms whose intermediates
fit into a fixed budget, and the two intermediate buffers are reused from chunk to chunk while
$\sigma$ accumulates. The device memory for intermediates is thus a parameter of the run and not
a function of $K$ and $D$.

The operands on the complementary side need one more step. Since the sites are assigned to the blocks in \cref{eq:decomposition}, the operators that enter a term are those of a block together
with its site. On the complementary side they are weighted sums of operators of the block
alone, of the stored ones and of the $P$ and $Q$ formed with \cref{eq:forming}. These sums are
not built on the host. The plan lists for each of them the operator blocks and weights it is
made of, and a kernel evaluates the sums on the device. If the resulting operators
fit into device memory, they are formed once and kept for all iterations of the bond. Otherwise
each chunk forms the ones it needs in a small window in every iteration, which trades time for
memory.

The Davidson iteration around the matrix--vector product also stays on the devices. The trial
vectors and their images under $H$ are held in device memory, each rank holding a part of every
vector, and the orthogonalization, the residual and the preconditioning with the diagonal of $H$
are computed there. Only the projected eigenproblem in the space of the trial vectors, a matrix
of a few tens of rows, is solved on the host. Apart from the diagonal of $H$, which is uploaded
once per bond, no object of the size of $\Psi$ crosses between host and device while a bond is
solved. A host implementation of the iteration remains as a fallback when device memory is
short.

If several states are requested, the iteration is a block Davidson method for the $k$ lowest
eigenvectors of the target sector. The matrix--vector product then applies $H$ to $k$ vectors in
one pass through the plan, and the decimation uses the weighted average of the density matrices
of the states.

\subsection{Renormalization}
\label{sec:impl-renorm}

The renormalization is the second expensive step. The decimation supplies the isometry $C$,
which maps the basis of the grown block, the $4D$ products of block states with site states,
onto the $D$ states that are kept, and every operator $O$ of the grown block has to be
transformed as $O' = C^{\dagger} O\, C$. The grown operators are sums of products of a stored
operator $O_n$ of the old block with a $4\times4$ site matrix $d$, with integrals in the
coefficients $w_n$ of $S$ and $H^{X}$. Building them explicitly would take sixteen times the
memory of the result. \code{} never builds them. $C$ is cut into four parts $C_a$, one for each
state $a$ of the site, the contributions that share a site matrix are summed on the device into
$F_d = \sum_n w_n O_n$, and the projection follows directly,
\begin{equation}
  O' = \sum_{d}\sum_{ab} d_{ab}\, C_a^{\dagger} F_d\, C_b^{\vphantom{\dagger}}.
  \label{eq:renorm}
\end{equation}
Forming and projection use the same batched GEMMs and reduction kernels as the matrix--vector
product and run in chunks bounded by a memory budget, so that only matrices of dimension $D$
exist at any time.

The renormalized $A$ and $B$ are written directly into a device buffer that stays allocated for
the next bond (\cref{fig:buffers}). There it is the source for the forming of $P$ and $Q$, for
the operands of the matrix--vector product and for the next renormalization, which writes into a
second buffer. The two buffers alternate along the sweep. As long as source and target fit into
device memory together, the pair operators of the growing block thus never leave the device
between the step that produces them and the steps that consume them. Where they do not fit, the
renormalized operators are moved to the host or streamed to disk in chunks, and they return to
the device at the next bond.

The decimation that precedes the renormalization diagonalizes the reduced density matrix sector
by sector. The sectors are divided among the ranks, large sectors are diagonalized on the
devices and small ones on the host, and the $D$ eigenvectors with the largest weights over all
sectors are kept.

\subsection{Memory hierarchy}
\label{sec:impl-memory}

The data of a calculation are spread over three tiers. Device memory holds the working set of
the current bond: the pair operators of the two blocks that the rank owns, the formed
complementary operators, the Krylov vectors and the intermediates of one chunk. Every transient
device buffer has an explicit budget, and the matrix--vector product and the renormalization
release their buffers to each other, so the peak is the larger of the two and not their sum.

The disk holds what belongs to the other bonds. A block is written in the background as soon as
it has been renormalized, so that the output overlaps with the computation. The operators of
the block on the other side of the bond, which the previous sweep wrote, are read back when the
bond is prepared.

Host memory is kept out of the leading terms as far as possible. Operators that are resident on
a device exist on the host only as their block structure and their diagonals, which the
Davidson preconditioner needs. Their values are streamed from the device to the disk in chunks
and come back for the opposite sweep direction, so that they pass through host memory only as a
transient. Besides this the host holds $H^{X}$, the plan and staging buffers for transfers. When
the renormalization of a bond does not fit on the device together with its sources, its result
is staged on the host, which then holds a transient of up to the size of the device memory.

\Cref{tab:operators} gives the resulting memory per rank. Which variant of a step is used is
decided bond by bond. Before a bond is prepared, the ranks agree on a memory plan. From the
block structure of the bond and the free device and host memory of the ranks, the plan selects
between the faster and the more economical variants of the steps above, for example whether the
complementary operands stay formed for all iterations and whether the renormalized pair
operators stay on the device. It records its prediction next to the measured peaks. After each
sweep the program also estimates the memory of the next sweep from the block structure of the
last one, and it warns, or on request stops, if a bond will not fit.

\subsection{Distribution over GPUs}
\label{sec:impl-mpi}

The terms of \cref{eq:decomposition} are independent of each other. Almost all of them are
labeled by a pair of orbitals, and so is most of the stored data. The parallelization
therefore distributes pairs~\cite{chan2004}. One map assigns every pair of spin orbitals
$(i\sigma, j\tau)$ to a rank. This rank stores $A_{ij}$ and $B_{ij}$ of every block that contains
both orbitals, renormalizes them, evaluates the terms of \cref{eq:decomposition} that contain
the pair and, in measurements, the density-matrix elements that need these operators. Storage
and work are therefore in the same place, and in the default mode no pair operator is ever sent
from one rank to another. The one-index operators $\cre{i\sigma}$ and $S_{i\sigma}$ are assigned
by spin orbital in the same manner. They are needed on more ranks than the one that stores them
and are exchanged between the ranks.

The map has to balance the work at every bond, although the blocks at different bonds contain
different subsets of the pairs. It is a table that is built once per run. For every block that
occurs in a sweep it balances the number of pairs per rank, separately for the spin
combinations $\alpha\alpha$, $\alpha\beta$ and $\beta\beta$ of a pair, whose terms differ in
cost, and the summed cost of the matrix--vector product including the terms with $S$.

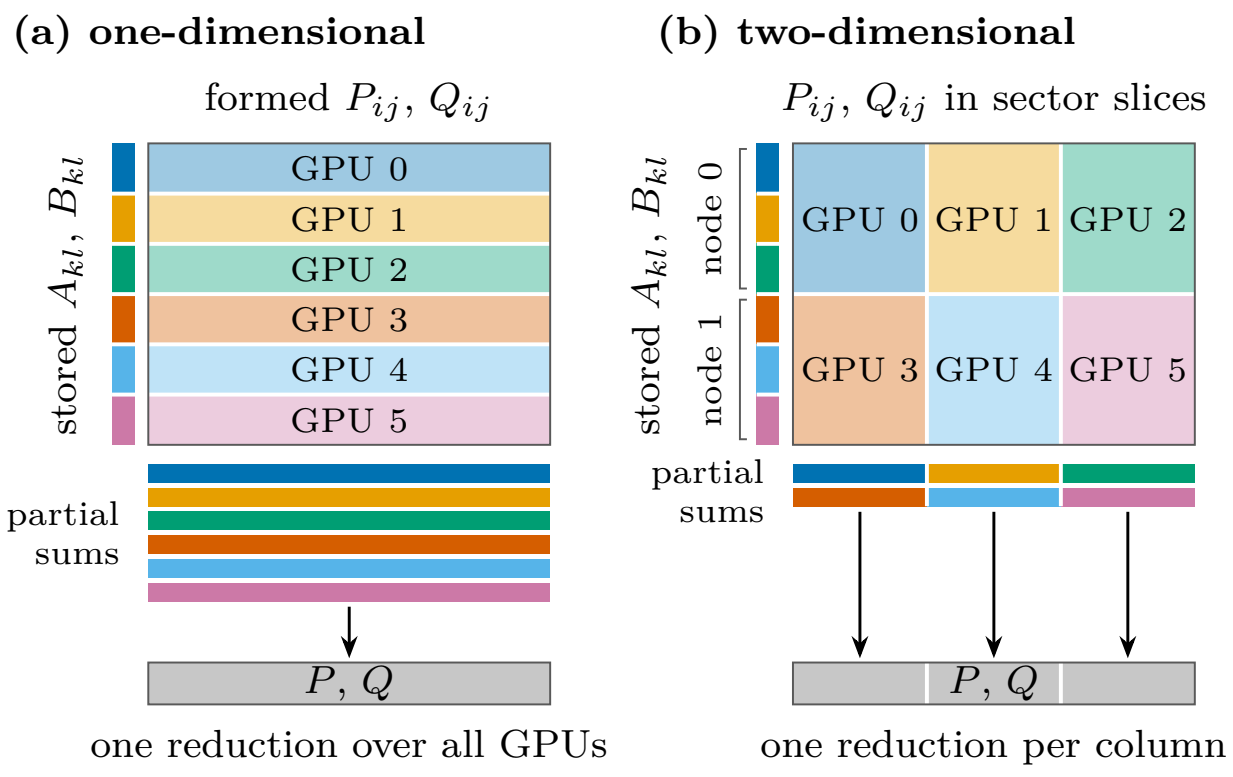
\begin{figure}[!hbt]
  \centering
  \resizebox{\columnwidth}{!}{\begin{tikzpicture}[font=\footnotesize, x=1cm, y=1cm, >={Stealth[length=1.6mm]}]
  \definecolor{g0}{RGB}{0,114,178}
  \definecolor{g1}{RGB}{230,159,0}
  \definecolor{g2}{RGB}{0,158,115}
  \definecolor{g3}{RGB}{213,94,0}
  \definecolor{g4}{RGB}{86,180,233}
  \definecolor{g5}{RGB}{204,121,167}
  \def\xl{0.98}    %
  \def\sqw{2.88}   %
  \def\sqh{2.16}   %
  \def\ybar{-2.30} %
  \def\yres{-3.72} %

  \begin{scope}
    \node[anchor=west, font=\footnotesize\bfseries] at (-0.12,0.78) {(a) one-dimensional};
    \node[anchor=south] at (\xl+0.5*\sqw,0.03) {formed $P_{ij}$, $Q_{ij}$};
    \node[rotate=90, anchor=south] at (\xl-0.34,-0.5*\sqh) {stored $A_{kl}$, $B_{kl}$};
    \foreach \i in {0,...,5} {
      \pgfmathsetmacro{\yt}{-\i*\sqh/6}
      \pgfmathsetmacro{\yb}{-(\i+1)*\sqh/6}
      \pgfmathsetmacro{\pt}{\ybar-0.17*\i}
      \fill[g\i!38] (\xl,\yb) rectangle (\xl+\sqw,\yt);
      \node[font=\scriptsize] at (\xl+0.5*\sqw,0.5*\yt+0.5*\yb) {GPU \i};
      \fill[g\i] (\xl-0.26,\yb) rectangle (\xl-0.10,\yt);
      \fill[g\i] (\xl,\pt-0.13) rectangle (\xl+\sqw,\pt);
    }
    \foreach \i in {1,...,5} {
      \draw[white, line width=0.8pt] (\xl-0.26,-\i*\sqh/6) -- (\xl+\sqw,-\i*\sqh/6);
    }
    \draw[black!65] (\xl,-\sqh) rectangle (\xl+\sqw,0);
    \node[anchor=east, align=right, font=\scriptsize] at (\xl-0.08,\ybar-0.49) {partial\\sums};
    \draw[->, semithick] (\xl+0.5*\sqw,\ybar-1.02) -- (\xl+0.5*\sqw,\yres+0.03);
    \fill[black!22] (\xl,\yres-0.30) rectangle (\xl+\sqw,\yres);
    \draw[black!65] (\xl,\yres-0.30) rectangle (\xl+\sqw,\yres);
    \node at (\xl+0.5*\sqw,\yres-0.15) {$P$, $Q$};
    \node[anchor=north] at (\xl+0.5*\sqw,\yres-0.36) {one reduction over all GPUs};
  \end{scope}

  \begin{scope}[xshift=4.62cm]
    \node[anchor=west, font=\footnotesize\bfseries] at (-0.12,0.78) {(b) two-dimensional};
    \node[anchor=south] at (\xl+0.5*\sqw,0.03) {$P_{ij}$, $Q_{ij}$ in sector slices};
    \node[rotate=90, anchor=south] at (\xl-0.78,-0.5*\sqh) {stored $A_{kl}$, $B_{kl}$};
    \foreach \i in {0,...,5} {
      \fill[g\i] (\xl-0.26,-\i*\sqh/6-\sqh/6) rectangle (\xl-0.10,-\i*\sqh/6);
    }
    \foreach \i in {1,2,4,5} {
      \draw[white, line width=0.8pt] (\xl-0.26,-\i*\sqh/6) -- (\xl-0.10,-\i*\sqh/6);
    }
    \foreach \r in {0,1} {
      \pgfmathsetmacro{\yt}{-\r*\sqh/2-0.04}
      \pgfmathsetmacro{\yb}{-(\r+1)*\sqh/2+0.04}
      \draw[black!65] (\xl-0.33,\yt) -- (\xl-0.38,\yt) -- (\xl-0.38,\yb) -- (\xl-0.33,\yb);
      \node[rotate=90, anchor=south, font=\scriptsize] at (\xl-0.39,0.5*\yt+0.5*\yb) {node \r};
    }
    \foreach \r in {0,1} {
      \foreach \c in {0,1,2} {
        \pgfmathtruncatemacro{\g}{3*\r+\c}
        \pgfmathsetmacro{\xa}{\xl+\c*\sqw/3}
        \pgfmathsetmacro{\xb}{\xl+(\c+1)*\sqw/3}
        \pgfmathsetmacro{\yt}{-\r*\sqh/2}
        \pgfmathsetmacro{\yb}{-(\r+1)*\sqh/2}
        \pgfmathsetmacro{\pt}{\ybar-0.17*\r}
        \fill[g\g!38] (\xa,\yb) rectangle (\xb,\yt);
        \node[font=\scriptsize] at (0.5*\xa+0.5*\xb,0.5*\yt+0.5*\yb) {GPU \g};
        \fill[g\g] (\xa,\pt-0.13) rectangle (\xb,\pt);
      }
    }
    \draw[white, line width=0.8pt] (\xl-0.26,-0.5*\sqh) -- (\xl+\sqw,-0.5*\sqh);
    \foreach \c in {1,2} {
      \draw[white, line width=0.8pt] (\xl+\c*\sqw/3,0) -- (\xl+\c*\sqw/3,\ybar-0.30);
    }
    \draw[black!65] (\xl,-\sqh) rectangle (\xl+\sqw,0);
    \node[anchor=east, align=right, font=\scriptsize] at (\xl-0.08,\ybar-0.19) {partial\\sums};
    \fill[black!22] (\xl,\yres-0.30) rectangle (\xl+\sqw,\yres);
    \foreach \c in {0,1,2} {
      \draw[->, semithick] (\xl+\c*\sqw/3+\sqw/6,\ybar-0.34) -- (\xl+\c*\sqw/3+\sqw/6,\yres+0.03);
    }
    \foreach \c in {1,2} {
      \draw[white, line width=0.8pt] (\xl+\c*\sqw/3,\yres) -- (\xl+\c*\sqw/3,\yres-0.30);
    }
    \draw[black!65] (\xl,\yres-0.30) rectangle (\xl+\sqw,\yres);
    \node at (\xl+0.5*\sqw,\yres-0.15) {$P$, $Q$};
    \node[anchor=north] at (\xl+0.5*\sqw,\yres-0.36) {one reduction per column};
  \end{scope}
\end{tikzpicture}}
  \caption{Forming of the complementary operators on six GPUs. Every formed operator
    (horizontal) is a sum over all stored pair operators (vertical), so the area of the
    rectangle is the work of the forming, and its colors show which GPU does which part. The
    bar at the left shows which pair operators a GPU stores, the bars below show the partial
    sums each GPU forms, which a reduction adds up. (a) Default mode: a GPU forms a partial sum
    of every complementary operator from its own pair operators. (b) Optional
    two-dimensional mode on two nodes with three GPUs each: every formed operator is cut along
    its symmetry sectors into three slices. A GPU forms one slice from the pair operators of
    its node, and a reduction over the nodes completes each slice. The partial sums are three
    times smaller.}
  \label{fig:distribution}
\end{figure}

What a rank cannot compute from its own operators are the sums over all pairs. These are the
complementary operators, for which the sum over $(kl)$ in \cref{eq:forming} runs over the pairs
of all ranks, and $\sigma$, which is the sum of the contributions of all terms. In the default,
one-dimensional mode every rank forms its partial sum of each of them
(\cref{fig:distribution}a), for example $\sigma^{(r)}$ from the terms of rank $r$, and the
partial sums are added, $\sigma = \sum_r \sigma^{(r)}$. For this the partial $P$, $Q$ and $S$ are
laid out identically on all ranks in a device buffer and completed by a reduction that runs from
device to device with NCCL~\cite{nccl}, after which every rank keeps the part it owns. The
partial $\sigma$ of a Davidson iteration and the $S$ and $H^{X}$ of a renormalized block are
completed in the same way. The ranks stay consistent because they run the same sweep, compile
identical plans from settings that the first rank broadcasts, and take every decision that
changes the control flow collectively.

This mode needs the least memory per rank, because every pair operator is held exactly once.
It has two limits. A rank forms and exchanges a partial sum of all complementary operators,
$O(K_N^2D^2)$ per bond, however many ranks there are, so this part of the work does not shrink
with $\nproc$. And a pair is the smallest unit of work, so with many ranks, each of which has
only a few pairs at a bond, the load can no longer be balanced well.

For calculations with Abelian symmetry an optional two-dimensional mode therefore distributes
along a second axis, the symmetry sectors (\cref{fig:distribution}b). The ranks form a grid of
$n_{\mathrm r}\times n_{\mathrm c}$, in which a row is by default the GPUs of one node and a
column contains one GPU of every node. The balanced pair map assigns the pairs to rows, and
every formed operator is cut into $n_{\mathrm c}$ slices, one per column. A slice is a set of
symmetry sectors, and a sector that is too heavy for one slice is divided into ranges of its
rows. A rank forms its slice of every $P$ and $Q$ from the pair operators of its row, which the
GPUs of a node exchange over the links within the node, and a reduction within the column
completes the slice. The partial sums formed and exchanged per rank shrink by the factor
$n_{\mathrm c}$. By default the operators $S$ are completed as in the
one-dimensional mode, and they can be completed by columns as well. With more columns than a
node has GPUs, a row consists of several nodes.

The mode has two levels. On the first only the forming is distributed in this way, and the
matrix--vector product runs as in the one-dimensional mode. On the second the matrix--vector
product is cut as well. The terms of a pair are evaluated on all ranks of its row, each with its
slice of the complementary operator. The slices are chosen by cost, with the work of every
sector counted at the last visit of the bond, so that the work of a row is divided nearly evenly
among its GPUs and the balance is that of a one-dimensional distribution over $n_{\mathrm r}$
ranks. In return every rank of a row needs the normal operators of the pairs of the row in full,
$n_{\mathrm c}$ times the memory of the normal side. A bond at which they do not fit into device
memory runs on the first level.

The balance of the terms with $S$ needs a separate treatment. There are only $2K$
operators $S_{i\sigma}$, so that with more GPUs than spin orbitals some ranks would evaluate the
terms of an $S_{i\sigma}$ and others none. In this case each $S_{i\sigma}$ is cut along its
symmetry sectors into parts, and a cost model places the parts on different ranks, the largest
part first and each on the rank with the least work so far. The parts act on different sectors,
so the split adds no operations. The working copy of the
$\cre{i\sigma}$ (\cref{tab:operators}) remains as the part of the memory per rank that does not
shrink with $\nproc$.

\subsection{Spin adaptation}
\label{sec:impl-su2}

For restricted orbitals \code{} can also exploit the SU(2) symmetry of the total
spin~\cite{sharma2012,wouters2014cpc}. The states are then labeled by particle number and total
spin, the operators are stored as reduced matrix elements of spin tensor operators, and $D$
counts multiplets. The coupling coefficients are numbers per pair of sectors and enter the
weights of \cref{eq:term,eq:renorm} at the place of the site matrix elements.
Spin-adapted calculations therefore run through the same device kernels, the same stores and
the same memory plan as Abelian ones. The balanced pair map and the split of $S$ apply to
them as well. Because the terms with $S$ carry a larger share of the work in spin-adapted
calculations, the split is used there at every number of ranks. Spin-adapted calculations use
the one-dimensional distribution.

Results that are defined through determinants are taken from the uncoupled form of a spin-adapted state. Each multiplet of spin $s$ on a bond is expanded with Clebsch--Gordan coefficients into its $2s+1$ states, and the state is represented by its member with $S_z = S$, which has U(1) symmetry. The RDMs, the orbital entanglement, the Slater determinants, the overlaps and the exported matrix product state use this form (\cref{sec:features}). Restarting a spin-adapted calculation requires its own saved bases, not the uncoupled state. The perturbative noise and the starting guess from configuration interaction are not yet available with SU(2) symmetry (\cref{sec:feat-aids}).

\subsection{Reproducibility and software}
\label{sec:impl-software}

Sums on a GPU are usually accumulated in an order that the thread scheduler sets. Results then
differ between runs in the last digits, and over the many Davidson iterations of a calculation
such differences can grow until they are visible in the energy. In \code{} every reduction
kernel gives each output element to exactly one thread, which adds the contributions in the
order the plan fixes, and concurrent streams are joined wherever two of them could write to the
same block. For a fixed number of ranks and fixed settings, two runs give energies and
discarded weights that agree in every digit. Runs with different numbers of ranks, or
on CPUs and on GPUs, add in a different order and can agree only to rounding. They do so as
long as the decimation is unambiguous. Where it has to choose among states of vanishing weight,
rounding decides which of them are kept, and from there on the runs follow different paths.
Converged calculations still agree to about $10^{-7}\,E_{\mathrm h}$ in our tests. Calculations
that are stopped before convergence, and calculations whose start lies close to an excited
state, can end in different places.

The core is written in C++20 with CUDA. All vendor-specific code, that is the kernels and the
calls into the BLAS, solver and communication libraries, sits behind one portability layer, so
that the support of another GPU vendor concerns this layer alone. The program is
driven from Python, directly or as the active-space solver of PySCF~\cite{sun2017,sun2020}
(\cref{sec:features}).
\section{Performance and scaling}
\label{sec:bench}

A machine with hundreds of GPUs helps a DMRG calculation only as far as a sweep scales with
their number. We measure this scaling in four series, chosen such that together they cover (i) both symmetries, (ii) weak and strong scaling, and (iii) active spaces
from small to large: the weak scaling of a small active space, P450(33,31) with SU(2) symmetry,
and of a large one, P-cluster(114,73) with U(1) symmetry, and the strong scaling of a medium
one, P450(51,50) with U(1) symmetry, and of a large one, FeMoco(113,76) with SU(2) symmetry. The
small weak-scaling and the medium strong-scaling series were also run on CPU nodes.
\Cref{fig:bench-main} shows the two large cases. The other series are described in the text,
and their figures can be found in the Supporting information (\cref{app:scaling}). All numbers were obtained with the release
version. The remainder of this section is about timings only. The runs were terminated after three sweeps, so their
energies are not converged and are not discussed. By contrast, converged energies are the subject of
\cref{sec:p450}.

\subsection{Setup}
\label{sec:bench-setup}

The GPU runs used the Booster module of JUPITER~\cite{jupiter}. A node has four GH200 superchips, each of them
an H100 GPU with 96\,GB of memory and a Grace CPU with 72 cores and 120\,GB of memory, and four
NDR200 InfiniBand links. Every MPI rank drives one GPU and uses the 72 cores of its Grace CPU. The
operators of the bonds that are not being solved are kept on the parallel scratch file system.
The program was built with CUDA 13.0 and uses NCCL 2.29.7 for the reductions between devices.

The CPU runs used the general-purpose partition of MareNostrum~5~\cite{marenostrum5}, whose nodes have Intel Xeon
Platinum 8480+ processors with 56 cores each, local NVMe storage and NDR200 InfiniBand. Every
MPI rank runs on one processor (socket) with its 56 cores and keeps the operators of the bonds that are not
being solved on the local NVMe storage.

Every point of every series follows the same protocol. A seed state is computed once per system
at a small bond dimension, 1000 states with U(1) and 800 multiplets with SU(2) symmetry, and
every run started from it. Three sweeps without noise followed. The first two grow the two blocks
to the target $D$. The third is the first sweep in which both sides of every bond have the full
$D$, and it is the one we report. Its wall time includes \textit{everything}: the forming of the
complementary operators, the matrix--vector products, decimation, renormalization and disk I/O.
The Davidson iteration stopped at a residual norm of $10^{-6}$, the default of the release, and
uses at most 40 vectors. Some programs compare the square of the residual norm, for which this
threshold is $10^{-12}$. All runs use the one-dimensional distribution (see Figure \ref{fig:distribution}), except for the
P-cluster, which was run in both modes.

Let $T(n)$ be the time of this sweep on $n$ ranks, $n = \nproc$. On the GPUs every rank drives
one GPU, so that $n$ is also the number of GPUs. In a strong-scaling series $D$ is fixed, and the efficiency is
\begin{equation}
  E(n) = \frac{\min_m\, m\,T(m)}{n\,T(n)},
\end{equation}
the cost of the sweep in GPU- or rank-hours relative to the most efficient point of the series.
In a weak-scaling series $D$ grows as $n^{1/3}$, which keeps the work per GPU constant if the
cost grows as $D^3$, as the leading $O(K^3D^3)$ does. The efficiency is then
\begin{equation}
  E(n) = \frac{\min_m\, T(m)}{T(n)},
\end{equation}
the shortest time per sweep of the series relative to that of the point. The exponent that is
actually found depends on the system. For P450(33,31) a single matrix--vector product grows as
$D^{2.9}$, and the Davidson method needs fewer iterations at larger $D$. The work of the
matrix--vector products of a sweep therefore grows as $D^{2.7}$, and the work per GPU falls
slowly along the series. For the P-cluster the number of iterations rises with $D$ instead,
and the work per GPU grows along the series (\cref{sec:bench-pcluster}).
In addition the GEMM rate grows with $D$, by which we mean the floating-point operations of
the GEMMs of the matrix--vector products per second and GPU. For P450(33,31) both effects make
the small points of the series the less efficient ones.

The figures divide the sweep into phases. The time of the matrix--vector products covers the
whole Davidson iteration and is split into three parts. Every application of $H$ ends with the
reduction of $\sigma$, which has to wait for the rank with the most work at the bond. The time
that the ranks wait there is the load imbalance, the time of the transfers is the
communication, and the rest is the computation. The other phases are the forming of the
complementary operators, the completion of the partial $P$, $Q$ and $S$ by the reduction between
the ranks (\cref{sec:impl-mpi}), the decimation, the renormalization and the prediction of the
starting vector. By \textit{useful operations} we denote the floating-point operations of
the GEMMs of the matrix--vector products and of the renormalization, with work that every rank
repeats counted once.

\subsection{Weak scaling of a small active space}
\label{sec:bench-weak}

Our small case is compound~I of P450 in the active space E of Goings et al.~\cite{goings2022},
P450(33,31), a doublet that is treated with SU(2) symmetry. The series starts on one GPU at
$D = 5200$, the largest of the tested values that fits into the memory of one device, and ends
on 16 GPUs at $D = 13\,100$ (\cref{fig:app-cpdie}).

Along the series the time per sweep does not stay constant, as ideal weak scaling would have
it, but falls to nearly half. Relative to the point on 16 GPUs, the efficiencies are 0.55,
0.68, 0.82 and 0.94 on 1, 2, 4 and 8 GPUs. The series thus scales better than ideal, and no
parallel overhead builds up along it. Communication stays below 1\,\%
of the sweep, and the matrix--vector products take about four fifths of the sweep at every
point. The gain on more GPUs comes from the two effects discussed in \cref{sec:bench-setup}. The GEMM
rate rises from 25 to 33\,TFLOPS per GPU, and the work grows more slowly than $D^3$.

\subsection{Strong scaling of a medium active space}
\label{sec:bench-strong}

The medium case is P450(51,50), the smallest of the active spaces of \cref{sec:p450-as} for
the reactant complex, with unrestricted orbitals and U(1) symmetry. The bond dimension is $D = 6900$, the largest of
the tested values that runs entirely in device memory on 8 GPUs, and the series goes from 8 to
64 GPUs (\cref{fig:app-p450}).

On eight times the GPUs the sweep becomes 4.7 times faster. Relative to 8 GPUs, the efficiency
is 0.90 on 16, 0.82 on 24, 0.75 on 32, 0.66 on 48 and 0.59 on 64 GPUs. The matrix--vector
products scale almost ideally over this range. They become 7.3 times faster, which is 91\,\%
of the ideal factor. The number of applications of $H$ per sweep and the GEMM rate are the same
at all points, so the loss is parallel overhead alone. It arises in the completion of $P$, $Q$ and $S$, which
takes about 40\,s per sweep at every point, because, as discussed in \cref{sec:impl-mpi}, in the one-dimensional mode every rank
receives its $O(K_N^2D^2)$ share of the completed operators however many GPUs there are available. Forming and completion together grow from 13 to 28\,\% of the sweep. At
this bond dimension the problem is small for 64 GPUs, and the fixed part dominates the loss
from 32 GPUs on.

\subsection{FeMoco with SU(2) symmetry}
\label{sec:bench-femoco}

The first large case is the FeMo cofactor, the active site of nitrogenase, in the model of Li
et al.~\cite{li2019femoco},
FeMoco(113,76) with total spin $S=3/2$, for which we use the publicly available Hamiltonian integrals of that work
(\cref{fig:bench-main}a, b). Several challenging DMRG calculations of this active space have been
published: with SU(2) symmetry at $D=4096$ on eight A100 GPUs~\cite{menczer2024b,menczer2024},
with unrestricted orbitals at $D=18\,000$ with Block2 on 24 CPU nodes with 2304
cores~\cite{zhai2026}, and with U(1) symmetry at $D=6000$ and at $D=3000$ on more than 2000
cores each~\cite{brabec2021,zhai2021}.
These DMRG calculations differ in hardware, symmetry and bond dimension, and we would like to stress that we quote them here for
orientation and not as a direct comparison.

Our series is a strong scaling at $D = 10\,000$ multiplets (about 80\,000 states), from 32 GPUs, the
smallest number on which the calculation fits, to 256 GPUs. From 32 to 256 GPUs the sweep
becomes 5.3 times faster, with efficiencies of 0.97 on 64, 0.83 on 128 and 0.66 on 256 GPUs
relative to 32 GPUs. The run on 256 GPUs \textbf{sustains 3.9\,PFLOPS of useful operations}.

FeMoco poses a harder problem for the Davidson method than the P450 active spaces. A two-site tensor optimization needs 30
applications of $H$ on average, against 8 for P450(51,50) (\cref{sec:bench-strong}) and 4 to 5
for P450(33,31) (\cref{sec:bench-weak}). The matrix--vector products therefore
take 97\,\% of the sweep on 32 GPUs and still 87\,\% on 256 GPUs. Their computation needs only
13\,\% more GPU-hours on 256 than on 32 GPUs, and the largest loss is load imbalance, which
stays at 6\,\% of the sweep up to 64 GPUs and grows to 10\,\% on 128 and 17\,\% on 256 GPUs.

\subsection{P-cluster with U(1) symmetry}
\label{sec:bench-pcluster}

Our second large case is the P-cluster, the [8Fe-7S] electron-transfer cofactor of
nitrogenase, in the model of Li et al.~\cite{li2019pcluster},
P-cluster(114,73), a singlet that is treated with U(1) symmetry (\cref{fig:bench-main}c, d).
\textsc{Focus} reached $D=14\,000$ for this active space on 48 A100 GPUs~\cite{xiang2024}. Our
series is a weak scaling from 64 to 512 GPUs in both modes of \cref{sec:impl-mpi}, from
$D = 14\,000$ to $28\,000$ in the one-dimensional mode and from $D = 12\,000$ to $24\,000$ in
the two-dimensional mode. Each series starts at a bond dimension at which the run on 64 GPUs
just fits into device memory with the fastest variant of every step (\cref{sec:impl-memory}).
The two-dimensional mode needs more device memory and therefore starts lower. Its runs use the
second level, on which the forming and the matrix--vector product are both cut by sectors, and
a row of the grid is the four GPUs of a node.

Over the whole series the two modes scale nearly alike. Relative to 64 GPUs, the efficiency is
0.92 on 128 and 0.87 on 256 GPUs in both modes, and on 512 GPUs it is 0.68 in the
one-dimensional and 0.66 in the two-dimensional mode. These efficiencies include a growth of
the work per GPU. The Davidson method needs 15 to 18\,\% more applications of $H$ per sweep at
the largest than at the smallest bond dimension of a series, so that a GPU performs 16\,\% more
useful operations per sweep on 512 than on 64 GPUs in the one-dimensional and 14\,\% more in
the two-dimensional mode. At equal work per GPU the efficiencies on 512 GPUs would be about
0.79 and 0.76. The largest run, the one-dimensional run
at $D = 28\,000$ on 512 GPUs, \textbf{sustains 9.6\,PFLOPS of useful operations} over the whole
sweep and 12.2\,PFLOPS during the matrix--vector products. For the same work a sweep of
the two-dimensional mode takes about 5 to 10\,\% longer than one of the one-dimensional mode, and at
the same $D$ it holds about 1.5 times the device memory, because every rank of a row needs the
normal operators of the pairs of the row.

The two modes differ in how far they can scale. What limits a distribution is the part of a
sweep whose cost per rank does not fall when ranks are added. In a weak scaling with
$D^3 \propto \nproc$ the work per rank stays the same, but other parts grow as $D^2$, that is
as $\nproc^{2/3}$. For example, the completion of $P$, $Q$ and $S$ behaves in this way. It grows nearly in
proportion to $D^2$, from 4\,\% of the sweep on 64 GPUs to 6\,\% on 128, 9\,\% on 256 and
10 to 11\,\% on 512 GPUs in both modes. Estimating this growth to 4096 GPUs, it would then be 16 times
as large as on 64 GPUs and lengthen the sweep execution to 1.6 times that on 64 GPUs. The efficiency
would then be about 0.6 even if every other part scaled ideally. In the
one-dimensional mode this is fixed by the layout, because every rank forms and receives a share
of every complementary operator. In the two-dimensional mode the part of $P$ and $Q$ is divided
by the number of columns $n_{\mathrm c}$ (\cref{sec:impl-mpi}). Already with the four columns
used here, and with $S$ completed as in the one-dimensional mode, the completion takes 6 to 14\,\% less time per $D^2$ than in the one-dimensional mode.

The same consideration holds for the balance. In the one-dimensional mode a pair is the smallest unit of
work, and the fewer pairs a GPU has at a bond, the less evenly they can be distributed. In the
two-dimensional mode the pairs are balanced over the rows, each of which has $n_{\mathrm c}$
times as many pairs as a single GPU would have, and the work of a row is divided among its GPUs
by sectors, in units far smaller than a pair. On 512 GPUs the load imbalance accordingly
takes 11\,\% of the sweep in the two-dimensional and 15\,\% in the one-dimensional mode. The balance therefore does not degrade in the
same way when GPUs are added, and more columns improve it further.

Hence, it is the two-dimensional mode that has the potential to enable a future scaling to thousands of GPUs. Here, we examined it in its initial conception. Its forming, which includes the exchange of the pair operators within a
node, still grows along the series, and the runs shown here used neither the completion of $S$
by columns nor more columns than a node has GPUs, both of which the program offers. None of
these limits are structural, and all of them can and will be removed in a future extension of this work to fully unlock the scaling of \code{} to thousands of GPUs.

Calculations of the size reported here also need large amounts of disk space. At the end of a sweep, the operators of the other bonds occupy up to 168\,TB in the largest run, summed over the
ranks, and about 90\,\% of it are the pair operators $A$ and $B$.

\subsection{CPU nodes}
\label{sec:bench-cpu}

Having discussed the scalings of the GPU framework, we now turn our attention to the CPU partition. Notably, \code{} runs on CPU nodes with the same algorithm and the same distribution, with
threaded host BLAS in place of the device kernels (\cref{sec:impl}). On MareNostrum~5 we
repeated two of the above series at smaller bond dimensions, the weak scaling of P450(33,31)
(\cref{fig:app-cpu-weak}) and the strong scaling of P450(51,50) (\cref{fig:app-cpu-strong}),
both on 1 to 16 ranks of 56 cores, that is on up to 896 cores. In the matrix--vector products a
rank of 56 cores reaches about a tenth of the rate of a GH200. For P450(51,50) at $D = 2500$ on
one rank they take 6490\,s per sweep on the CPU and 676\,s on a GPU, for the same work. The
weak-scaling series reaches efficiencies of 0.76, 0.89, 0.97 and 0.97 on 1, 2, 4 and 16 ranks,
relative to the point on 8 ranks. The strong-scaling series becomes 12.2 times faster on 16
times the ranks, with efficiencies of 0.98, 0.95, 0.96 and 0.75 on 1, 4, 8 and 16 ranks,
relative to 2 ranks.

In strong scaling the CPU series scales better than the GPU series. On eight times the ranks,
from 2 to 16, the sweep becomes 6.0 times faster, with an efficiency of 0.75, against 4.7 times
and 0.59 on eight times the GPUs. The forming and the renormalization also
scale with the number of ranks: they become 8.4 and 7.9 times faster from 2 to 16 ranks. The
completion of $P$, $Q$ and $S$ is again a fixed time per sweep, 160 to 220\,s on 2 to 16 ranks,
and causes most of the loss on 16 ranks, where it takes 21\,\% of the sweep. In the weak-scaling
series it grows from 32\,s on 2 ranks to 197\,s, 7\,\% of the sweep, on 16 ranks, which makes the
point on 16 ranks slower than that on 8. Communication and load imbalance each stay below
2\,\% of the sweep.

\subsection{Scaling analysis}
\label{sec:bench-analysis}

In summary, the series discussed in the previous section showed that the release version of \code{} scales well into the hundreds of GPUs and, on CPU nodes, to thousands of cores (shown here up to 896). In the following, we will take a closer look which parts of a sweep carry this scaling and which limit it.

The scaling is carried by the matrix--vector products, the largest part of the sweep in all
cases. Their GEMM work is distributed over the pair operators, and the only exchange between the
ranks that they need is the reduction of $\sigma$ once per application. Communication
accordingly takes at most 3\,\% of the sweep in every series, and in the strong-scaling series of
P450(51,50) the matrix--vector products keep 91\,\% of the ideal speed-up on eight times the GPUs. The GPUs
themselves are used well. The GEMMs of the matrix--vector products sustain up to 34\,TFLOPS per
GPU, about half of the theoretical double-precision peak of an H100
with tensor cores, and the rate grows with the bond dimension (\cref{sec:bench-weak}), so the
large calculations are the efficient ones.

Three smaller parts limit the scaling of the present release version. The first is the completion of
$P$, $Q$ and $S$ in the one-dimensional mode, a fixed $O(K_N^2D^2)$ per rank
(\cref{sec:bench-strong}). The second is the load imbalance of the matrix--vector product,
which grows as the pair operators per GPU become few. The third are the parts of
the forming and of the renormalization that every rank does once per bond. The two-dimensional
mode addresses the first two. It balances the pairs over the rows of its grid, and divides
the part of the completion that it distributes, that is, $P$ and $Q$, by the number of its
columns (\cref{sec:bench-pcluster}).

Forming $P$ and $Q$ only at the bond that in practive needs them (see \cref{sec:impl-operators}) costs little
time compared to the saving it entails. The forming with its completion takes about a tenth of the sweep or less
up to 128 GPUs in the two weak-scaling series. For the P-cluster on 512 GPUs it reaches 14\,\% in the one-dimensional and about a fifth in
the two-dimensional mode, and it reaches 28\,\% at the end of the strong-scaling series of
P450(51,50), where the problem is small compared to the number of GPUs. For FeMoco it stays below 6\,\% up to 256 GPUs. The forming is
done once per bond and the matrix--vector product once per Davidson iteration, so the share of
the forming grows when fewer iterations are needed. In return, considering for example the P-cluster at $D = 28\,000$ on 512 GPUs, about 34\,TB of $P$ and $Q$
per sweep are never written to or read from disk.

\begin{figure*}[!htb]
  \centering
  \includegraphics[width=\textwidth]{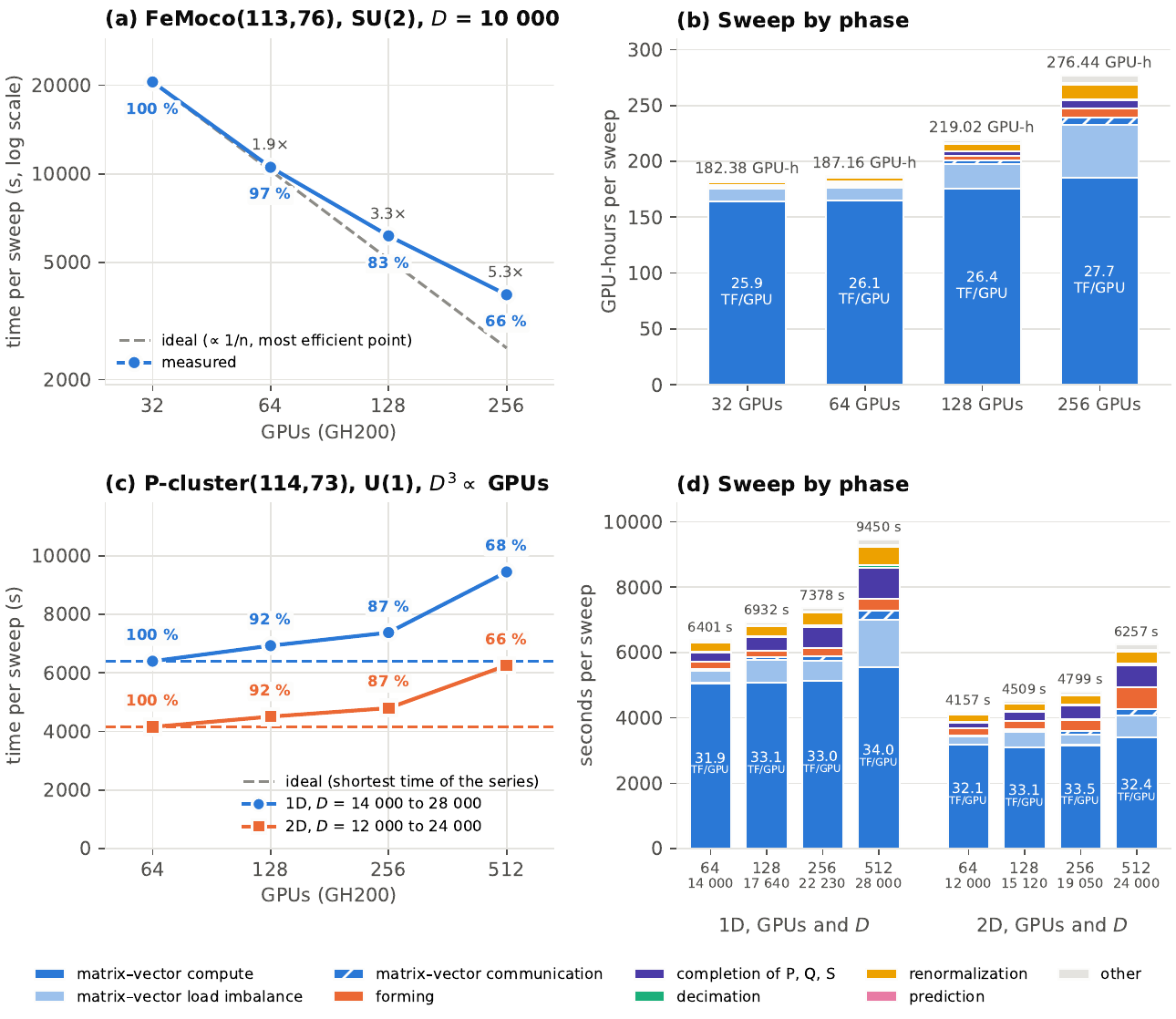}
  \caption{The two large cases. (a), (b) Strong scaling of FeMoco(113,76) with SU(2),
    $D = 10\,000$ on 32 to 256 GPUs. (c), (d) Weak scaling of P-cluster(114,73) with
    U(1), $D^3 \propto n$, on 64 to 512 GPUs, from $D = 14\,000$ to 28\,000 in the
    one-dimensional and from 12\,000 to 24\,000 in the two-dimensional mode. (a), (c) Time per sweep with the
    efficiency at every point, relative to the most efficient point of the series, in (a) also
    with the speed-up over 32 GPUs. Dashed: ideal scaling through the most efficient point. (b), (d) The
    sweep by phase, (b) in GPU-hours (the number of GPUs times the time, level bars for ideal
    strong scaling), (d) in seconds. The number in a bar is the GEMM rate per GPU of the
    matrix--vector products.}
  \label{fig:bench-main}
\end{figure*}
\section{Application: Drug metabolism mediated by cytochrome P450}
\label{sec:p450}

The two cofactors of \cref{sec:bench} have been studied with large DMRG calculations many
times, most recently by Chan and co-workers \cite{zhai2026}. As an application we therefore chose to look closer at a system of pharmaceutical interest instead, namely the initial metabolic breakdown of a drug molecule mediated by 
cytochrome P450. The enzymes of this family catalyze the oxidative metabolism of more than
three quarters of the drugs in clinical use~\cite{guengerich2008}. One of them, CYP2C9, breaks
down the anticoagulant warfarin, a drug with a narrow therapeutic window, mainly by
hydroxylation of its (S)-enantiomer at the carbon atom C7~\cite{civic2026}. How fast a drug is
metabolized follows from the barrier of the rate-determining step, which here is the attack of
the reactive intermediate of the enzyme, compound~I, on the aromatic ring of warfarin. The rate
depends exponentially on this barrier, so the energies of the reactant complex (RC) and of the
transition state (TS) have to be known to about 1\,kcal/mol.

Compound~I is an iron(IV)--oxo porphyrin radical with three unpaired electrons and close-lying
doublet and quartet states, and barriers from density functional theory depend strongly on the
functional~\cite{civic2026}. This makes the reaction a suitable target for multiconfigurational DMRG calculations. Earlier DMRG studies of
P450 treated models of compound~I without a substrate, in active spaces of up to 43 orbitals at
$D=1500$~\cite{goings2022} and of 58 orbitals with spin adaptation at
$D\le4096$~\cite{menczer2024}. The chemistry of the reaction, from the model of the enzyme
with its substrate to the comparison with experiment, is the subject of a separate
paper~\cite{civic2026}. This section is about the DMRG calculations: the active spaces, the
protocol, and the energies of RC and TS.

\subsection{Active spaces}
\label{sec:p450-as}

The structures of RC and TS come from QM/MM calculations of the solvated
enzyme~\cite{civic2026}. The quantum region contains the porphyrin with the iron--oxo unit, the
axial thiolate ligand and warfarin in its deprotonated form, and the protein enters as point
charges. Both structures are described as doublets with unrestricted Hartree--Fock in the
cc-pVDZ basis.

The active spaces are selected with AEGISS~\cite{aegiss}. The occupied $\alpha$ and $\beta$
orbitals of each structure are localized separately with the Pipek--Mezey
method~\cite{pipek1989}, and the quantum region is divided into chemically meaningful
fragments, such as the iron--oxo unit, the porphyrin ring, the thiolate and the rings of
warfarin. For each fragment the localized orbitals are projected onto its atomic orbitals,
those with the largest contributions are selected, and every selection is checked visually.
The virtual orbitals are frozen natural orbitals~\cite{taube2005,taube2008} tailored to this
selection. An MP2 calculation correlates only the selected occupied orbitals, and its
one-particle density matrix in the virtual space is diagonalized separately for $\alpha$ and
$\beta$. For each spin, the natural orbitals with the largest occupations are added until the
active space has the intended number of orbitals, which is the same for both spins.

We use three active spaces. The smallest, P450(51,50), contains the Fe 3d and oxo 2p orbitals,
the $\pi$ systems of the porphyrin and of the benzo ring of warfarin, the $\sigma$ lone pairs
of the porphyrin nitrogen atoms and one 3p lone pair of sulfur. P450(77,76) adds the remaining
3s and 3p lone pairs of sulfur, the $\pi$ system of the pyranone ring, which completes that of
the warfarin ring system, and the $\sigma$ bonds of the benzo ring together with the C7--H bond
at the site of the attack. The number of occupied $\alpha$ and $\beta$ orbitals grows from 26
and 25 to 39 and 38 for both structures. P450(77,100) adds 24 further natural orbitals per spin
to recover more of the correlation.

For the DMRG calculations the orbitals of the two spins are then paired on the sites: within
the occupied and within the virtual space, corresponding orbitals combine the $\alpha$ and
$\beta$ orbitals into pairs of largest overlap, the one unpaired occupied $\alpha$ and the one
unpaired virtual $\beta$ orbital share a site, and each block of pairs is localized with one
Pipek--Mezey rotation that is applied to both spins. This raises the mean overlap of the two
orbitals of a site from below 0.15 to about 0.8 and leaves the Hartree--Fock determinant and
the exact energies in the active space unchanged.

\subsection{DMRG calculations}
\label{sec:p450-protocol}

For each active space we compute the ground states of RC and TS with U(1) symmetry and
spin-resolved integrals (\cref{sec:impl-algorithm}), in the sector with $2S_z = 1$, on the
GH200 nodes of JUPITER (\cref{sec:bench-setup}). The six calculations follow the same protocol.
The orbitals are ordered by the Fiedler vector of the exchange integrals $|(pq|qp)|$
(\cref{sec:feat-aids}). A calculation starts from the Hartree--Fock determinant with 20 sweeps
at $D=1000$, of which the first six carry noise (\cref{sec:feat-aids}) with an amplitude of
$10^{-4}$ and the next four with $10^{-5}$, to keep it from settling in a local minimum. The
bond dimension is then raised to its maximum, $D_{\max} = 7000$, 9000 and 12\,000 for the three
active spaces, with noise in the first four sweeps, two at $10^{-4}$ and two at $10^{-5}$.
Sweeps without noise follow until the energy improves by less than $0.1\,\mathrm{m}E_{\mathrm h}$
over four sweeps. In P450(51,50) the calculation at $D_{\max}$ took 14 sweeps for RC and 14
for TS, respectively. The Davidson iteration stops at a residual norm
of $2\times10^{-5}$, and of $10^{-6}$ in the last two sweeps.

The energy of the exact ground state in the active space is estimated by an extrapolation in
the discarded weight, the weight of the density matrix that the decimation drops
(\cref{sec:impl-algorithm})~\cite{legeza1996,olivares2015}. Starting from the converged state,
the bond dimension is lowered in five steps to about two thirds of $D_{\max}$. At every step the state of the previous one is optimized
with six sweeps without noise. Each step gives one point, its energy with the largest discarded
weight of its last sweep. A linear fit of the energy against the discarded weight gives the
extrapolated energy $E_\infty$, and quadratic fits and fits to subsets of the points give an
estimate of its uncertainty.

\subsection{Results}
\label{sec:p450-results}

\begin{table}[!htb]
  \caption{DMRG energies of the reactant complex (RC) and of the transition state (TS) of the
    hydroxylation of warfarin by CYP2C9 in the active space P450(51,50): the lowest variational
    energy
    $E(D_{\max})$ and the energy $E_\infty$ extrapolated to zero discarded weight with its
    uncertainty, both in $E_{\mathrm h}$. $\Delta E$ is the
    difference between TS and RC in kcal/mol.}
  \label{tab:p450}
  \centering
  \footnotesize
  \vspace{3pt}
  \renewcommand{\arraystretch}{1.15}
  \begin{tabular}{@{}l@{\hspace{6pt}}r@{\hspace{6pt}}l@{\hspace{6pt}}l@{\hspace{6pt}}l@{}}
    \toprule
    Active space & $D_{\max}$ & & $E(D_{\max})$ & $E_\infty$ \\
    \midrule
    P450(51,50)  & 7000     & RC         & $-3785.20489$ & $-3785.2066 \pm 0.0013$ \\
                 &          & TS         & $-3785.14453$ & $-3785.1453 \pm 0.0004$ \\
                 &          & $\Delta E$ & 37.87 & $38.5 \pm 0.6$ \\

    \bottomrule
  \end{tabular}
\end{table}

\Cref{tab:p450} collects the energies for P450(51,50). At $D_{\max} = 7000$ the discarded
weights are $2.5\times10^{-5}$ for RC and $1.7\times10^{-5}$ for TS, and the variational energies
lie 1.7\,m$E_{\mathrm h}$ (RC) and 0.8\,m$E_{\mathrm h}$ (TS) above the extrapolated ones. The
difference between TS and RC is 37.87\,kcal/mol at $D_{\max}$ and 38.5\,kcal/mol in the
extrapolation, with an uncertainty of 0.6\,kcal/mol from the extrapolation. Each of the two
calculations ran on 32 GPUs, and a sweep at $D_{\max}$ took about six minutes.

Notably, as will be shown in Ref.~\cite{civic2026}, this difference is too large compared to experimental evidence for the barrier of the reaction and points to a lack of dynamical correlation that cannot be recovered even within large(r) active-space Hamiltonians. To this end, we will put forward a simple composite correction scheme in Ref.~\cite{civic2026} which allows us to estimate dynamical correlation contributions from large-scale coupled-cluster calculations.

Notably, the calculations in the two larger active spaces, P450(77,76) at $D_{\max} = 9000$ and
P450(77,100) at $D_{\max} = 12\,000$, have been started but could not be completed for this
version, because of repeated interruptions of the operation of JUPITER and a longer scheduled
maintenance. Their results will be added in an updated version of this paper as soon as the
machine is back in regular operation.

\section{Further capabilities}
\label{sec:features}

Ground-state energy calculations are rarely the end of a quantum-chemistry workflow. In this section, we will highlight the extended capabilities \code{} offers, in the order of a regular quantum-chemistry workflow: input, solution, measurement, analysis
and orbital optimization. \Cref{tab:features} lists the capabilities at the time of release and
states for each one with which symmetry it is available and whether it runs on GPUs and
distributed over several MPI ranks.

All examples of this section use one system, the iron--sulfur cluster
[Fe$_2$S$_2$(SCH$_3$)$_4$]$^{2-}$~\cite{sharma2014} in an active space of 30 electrons in 20
orbitals. Its two high-spin Fe(III) centers couple to a ladder of states with total spin $S=0$
to $5$. This makes it a small relative of the systems of \cref{sec:bench,sec:p450}, and every
calculation of this section runs on one GPU. A spin-adapted state at $D=2000$ takes one to seven
minutes, the six states of \cref{sec:feat-excited} with U(1) symmetry 82 minutes and the two
stages of the CASSCF example 101 minutes.
The integrals of the active space are
public~\cite{li2017,fe2s2data}, together with the energy of the singlet ground state in the
active space, $-116.605\,609\,1\,E_{\mathrm h}$~\cite{fe2s2data}.

\begin{table}[!htb]
  \caption{Capabilities of \code{} at the time of release, with U(1) and with SU(2) symmetry,
    on GPUs and distributed over several MPI ranks. A dash marks a step that runs on the host before or after
    the sweeps. Orbital entanglement stands for the single-orbital entropies and the mutual
    information.}
  \label{tab:features}
  \centering
  \footnotesize
  \vspace{3pt}
  \renewcommand{\arraystretch}{1.12}
  \newcommand{\yes}{$\checkmark$}
  \begin{tabular}{@{}l@{\hspace{6pt}}c@{\hspace{7pt}}c@{\hspace{7pt}}c@{\hspace{7pt}}c@{\hspace{6pt}}l@{}}
    \toprule
    Capability & U(1) & SU(2) & GPU & MPI & Section \\
    \midrule
    Ground state & \yes & \yes & \yes & \yes & \ref{sec:bench}, \ref{sec:p450} \\
    Excited states & \yes & \yes & \yes & \yes & \ref{sec:feat-excited} \\
    1- and 2-particle RDMs & \yes & \yes\textsuperscript{a} & \yes & \yes & \ref{sec:feat-measure} \\
    $\langle S^2\rangle$ & \yes & \yes\textsuperscript{a} & \yes & \yes & \ref{sec:feat-measure} \\
    Orbital entanglement\textsuperscript{b} & \yes & \yes\textsuperscript{a} & \yes & \yes & \ref{sec:feat-measure} \\
    Slater determinants & \yes & \yes\textsuperscript{a} & \yes & \yes & \ref{sec:feat-dets} \\
    CASSCF & \yes & \yes & \yes & \yes & \ref{sec:feat-casscf} \\
    Order from integrals & \yes & \yes & -- & -- & \ref{sec:feat-aids} \\
    Order from entanglement & \yes & \yes & -- & -- & \ref{sec:feat-aids} \\
    CI starting guess & \yes & & -- & -- & \ref{sec:feat-aids} \\
    Perturbative noise & \yes & & \yes & \yes & \ref{sec:feat-aids} \\
    Checkpoint and restart & \yes & \yes & \yes & \yes & \ref{sec:feat-aids} \\
    Export, overlaps & \yes & \yes\textsuperscript{a} & -- & -- & \ref{sec:feat-aids} \\
    \bottomrule
  \end{tabular}

  \vspace{2pt}
  \raggedright\footnotesize
  \textsuperscript{a}\,Through the uncoupled U(1) form of the state.
  \textsuperscript{b}\,For the lowest state when several states are computed.
\end{table}

\subsection{Python and PySCF interface}
\label{sec:feat-interface}

\code{} is driven from Python. A calculation is held by a session object. Its constructor takes
the integrals, the symmetry, the target sector and the resources: the GPU, the MPI communicator, the number of threads and whether operators are kept out of core. One call then computes the state for a given bond dimension and number of sweeps. The state stays with the session, in memory or on
disk, and all later steps work on it: measurements, a change of the bond dimension for an
extrapolation, a restart with new integrals and the export of the matrix product state. The
integrals are passed as arrays, for example from an FCIDUMP file, so the solver itself does not
depend on a quantum chemistry package.

For calculations that start from a molecule, \code{} provides an active-space solver for
PySCF~\cite{sun2017,sun2020}. It takes the place of the FCI solver in CASCI and CASSCF calculations
with restricted and with unrestricted orbitals, and PySCF obtains energies, density matrices
and $\langle S^2\rangle$ from it through its usual interface. The target spin follows the convention of
PySCF and is taken from the numbers of $\alpha$ and $\beta$ electrons in the active space,
$2S = N_\alpha - N_\beta$. \Cref{lst:casscf} shows a
DMRG-CASSCF calculation, the second stage of the example of \cref{sec:feat-casscf}. The same
script runs on several GPUs when it is started under MPI with one rank per GPU. All settings are keyword arguments whose defaults need no tuning.

\begin{lstlisting}[float=t, caption={DMRG-CASSCF calculation with PySCF: the second stage
  of the example of \cref{sec:feat-casscf}.},
  label=lst:casscf]
from pyscf import gto, lib, scf
from exa_dmrg import DMRGSCF

mol = gto.M(atom="fe2s2.xyz", charge=-2,
            basis="tzp-dkh")
mf = scf.RHF(mol).sfx2c1e()   # the Hamiltonian
mf.mo_coeff = lib.chkfile.load(
    "stage1.chk", "mcscf/mo_coeff")

mc = DMRGSCF(mf, 20, 30, max_bond=1000,
             solver_kwargs={"sym": "su2"})
mc.fcisolver.mps_restart = True  # seeded restarts
mc.conv_tol, mc.conv_tol_grad = 1e-6, 0.1
mc.chkfile = "stage2.chk"     # orbitals
mc.kernel()
# several GPUs: mpirun -n 4 python casscf.py
\end{lstlisting}

\subsection{Excited states}
\label{sec:feat-excited}

Several states of one symmetry sector are computed in a state-averaged calculation. Up to ten
states share one set of renormalized bases, which the decimation selects from the weighted
average of their density matrices. The weights are equal unless the user sets them to be otherwise. At each
bond the block Davidson method of \cref{sec:impl-matvec} solves for all states together, and
one pass of the matrix--vector product serves several of them. In the example at $D=1000$, a
sweep with U(1) symmetry takes 2.0, 3.6 and 7.5 times as long for 3, 6 and 10 states as for one
state. With SU(2) symmetry a sweep for three states takes 2.8 times as long as for one. Energies, reduced density matrices (RDMs) and $\langle S^2\rangle$ are available
for every state, and the state-averaged RDMs are formed in the engine. A restart can change the
number of states.

What a sector contains differs between the two symmetries. With SU(2) symmetry all states of a
calculation have the same total spin. With U(1) symmetry a sector with spin projection $S_z$
contains the states of every $S \ge |S_z|$, so states of different spin come out of one
calculation and are told apart by their $\langle S^2\rangle$. In both cases the result depends
on where the list of states ends. If the state above the last requested one is nearly
degenerate with it, the averaged density matrix depends on which of the two the solver has
kept. \code{} therefore solves once for one state more at the end of a calculation and warns if
that state lies within $10^{-4}\,E_{\mathrm h}$ of the last one.

\begin{figure}[!htb]
  \centering
  \includegraphics[width=\columnwidth]{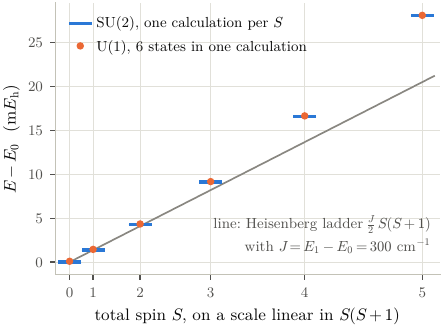}
  \caption{Spin ladder of the [2Fe-2S] cluster. The bars are the lowest state of each total
    spin from six spin-adapted calculations at $D=2000$, relative to the singlet. The points
    are the six states of one state-averaged calculation with U(1) symmetry at
    $D=2000$, placed at their $\langle S^2\rangle$. The line is the ladder of a Heisenberg model
    with the exchange coupling taken from the singlet--triplet gap.}
  \label{fig:ladder}
\end{figure}

The spin ladder of the [2Fe-2S] cluster illustrates both routes (\cref{fig:ladder}). With SU(2)
symmetry, six calculations at $D=2000$ give the lowest state of each spin from $S=0$ to $5$.
The singlet lies $3\times10^{-7}\,E_{\mathrm h}$ above the reference energy. The triplet follows
$1.37\,\mathrm{m}E_{\mathrm h}$ above the singlet. In a Heisenberg model of the two iron spins,
$E(S) = E_0 + \tfrac{J}{2}S(S+1)$, this gap is the exchange coupling, $J = 300\,\mathrm{cm}^{-1}$.
The states of higher spin lie above the ladder of this model, by 4\,\% for $S=2$ and by 37\,\%
for $S=5$. With U(1)
symmetry, one state-averaged calculation of six states in the sector $S_z=0$ at $D=2000$ gives
the same ladder. The six lowest states of the sector have $\langle S^2\rangle = 0.001$, 2.000,
5.999, 12.000, 19.999 and 29.999, that is $S(S+1)$ for $S=0$ to $5$, and their energies lie
between $2\times10^{-5}$ and $9\times10^{-5}\,E_{\mathrm h}$ above those of the spin-adapted
calculations.

The ladder also shows what the choice of the symmetry means for a truncated calculation. With
SU(2) symmetry the total spin of a state is fixed by construction at every bond dimension. With
U(1) symmetry only $S_z$ is fixed, and the calculation finds the spin of a state only as far as
it resolves the states of the sector. In this cluster they lie close together. At $D=250$ the
error of the ground-state energy is 3\,m$E_{\mathrm h}$ (\cref{fig:order}), twice the gap
between singlet and triplet, and the state that a calculation with U(1) symmetry finds has
$\langle S^2\rangle = 2.0$ instead of 0. At $D=100$ it has 3.9. A state that is optimized at
$D=2000$ and then projected to smaller bond dimensions stays a singlet down to $D=500$, with
$\langle S^2\rangle$ below 0.01. The examples of \cref{sec:feat-dets,sec:feat-aids} meet this
mixing of spin states again.

\subsection{Measurements}
\label{sec:feat-measure}

Measurements are taken on the state that a calculation has retained. A separate pass over the
chain, in which nothing is optimized or truncated, brings the state into canonical form and
evaluates the one- and two-particle RDMs element by element on the GPUs. The elements are
distributed over the ranks like the operators they are built from (\cref{sec:impl-mpi}). The
RDMs therefore all belong to one state, the matrix product state that the calculation has
retained. A direct test is the energy recomputed from them,
\begin{equation}
  E = \sum_{pq} h_{pq}\,\gamma_{pq} + \frac12 \sum_{pqrs} (pq|rs)\,\Gamma_{pqrs}.
\end{equation}
For the ground state of the example at $D=2000$ with U(1) symmetry it differs from the energy
of the calculation by $7\times10^{-6}\,E_{\mathrm h}$, at a discarded weight of
$6\times10^{-6}$. The difference is the truncation of the last decimation: the energy of the
calculation is that of the two-site wave function before it, and the RDMs are those of the
stored state after it. The measurement is cheap. The pass for the one- and two-particle RDMs
takes 18\,s on one GPU, less than one sweep of the calculation, which takes 29\,s. The RDMs are
returned spin-resolved or spin-traced. For spin-adapted states
they are evaluated on the uncoupled U(1) form of the state.

$\langle S^2\rangle$ is available for every state. For spin-adapted states the symmetry fixes
it at $S(S+1)$, and the program measures it on the uncoupled form like the RDMs. With
unrestricted orbitals the $\alpha$ and
$\beta$ orbitals differ in space, and the spin of the active space then follows from the RDMs
together with the overlap of the two sets of orbitals.

\begin{figure}[!htb]
  \centering
  \includegraphics[width=\columnwidth]{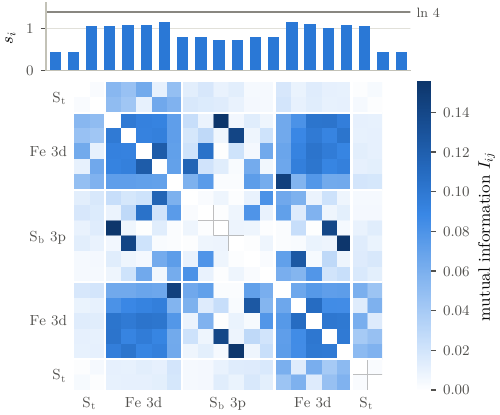}
  \caption{Orbital entanglement in the ground state of the [2Fe-2S] cluster ($D=2000$, U(1)
    symmetry). Top: single-orbital entropies $s_i$, with their maximum $\ln 4$. Bottom: mutual
    information $I_{ij}$ of all pairs of orbitals. The orbitals are in the order of the
    published integrals: 3p orbitals of the terminal thiolate sulfur atoms (S$_\mathrm{t}$), the
    3d orbitals of the two iron atoms, and between them the 3p orbitals of the bridging sulfur
    atoms (S$_\mathrm{b}$).}
  \label{fig:mi}
\end{figure}

The single-orbital entropies and the mutual information of pairs of
orbitals~\cite{legeza2003,rissler2006} show how strongly an orbital is entangled with the rest
of the active space and with which other orbitals. They help to choose an active space and to
order the orbitals (\cref{sec:feat-aids}). \Cref{fig:mi} shows both for the ground state of
the [2Fe-2S] cluster. The ten Fe 3d orbitals carry entropies of 1.0 to 1.2, not far from the
maximum of $\ln 4$, as expected for two high-spin centers. The 3p orbitals of the bridging
sulfur atoms follow with 0.7 to 0.8, and the orbitals of the terminal thiolate groups are the
least entangled. The largest values of the mutual information connect a Fe 3d orbital with a 3p
orbital of a bridging sulfur atom, and the 3d orbitals of the two centers with each other.

\subsection{Slater determinants}
\label{sec:feat-dets}

A matrix product state gives direct access to the coefficient of any Slater determinant. The
occupation of each orbital selects one matrix per site, and the coefficient is the product of
these matrices. \code{} evaluates the coefficients of a given list of determinants, and it
searches for all determinants whose coefficient exceeds a threshold in
magnitude~\cite{lee2021,zhai2023}. The search extends partial occupations orbital by orbital.
In right-canonical form the norm of the partial product bounds the coefficient of every determinant
that begins with it. A partial occupation is therefore dropped as soon as this norm falls below
the threshold, and no determinant above the threshold is missed. All surviving partial occupations of a step are advanced together as batched matrix products,
on the host or on a GPU. On a GPU, a step whose partial occupations do not fit into the device
memory is split into parts that are completed one after the other.

\begin{figure}[!htb]
  \centering
  \resizebox{0.80\columnwidth}{!}{\begin{tikzpicture}[font=\footnotesize, x=1cm, y=1cm, >={Stealth[length=1.6mm]}]
  \definecolor{g0}{RGB}{0,114,178}
  \definecolor{g1}{RGB}{230,159,0}
  \definecolor{g2}{RGB}{0,158,115}
  \def\dy{0.7}     % one orbital
  \def\yK{-4.0}    % last orbital
  \def\ybar{-4.42} % lists of the ranks
  \def\yres{-5.05} % merged result
  \def\xs{0.2}     % half spacing of the spine rows
  \def\xo{-4.1}    % orbital axis
  \tikzset{occ/.style={fill=white, inner sep=1pt, font=\scriptsize},
           edge/.style={black!45}, deal/.style={->, black!60, shorten >=1.2mm}}

  % orbital axis
  \node[anchor=east] at (\xo,2.2) {orbital};
  \node[anchor=east] at (\xo,1.05) {1};
  \node[anchor=east] at (\xo,0.55) {$\vdots$};
  \node[anchor=east] at (\xo,0) {$k$};
  \foreach \k in {1,...,4} {
    \node[anchor=east] at (\xo,-\k*\dy) {$k{+}\k$};
  }
  \node[anchor=east] at (\xo,\yK+0.55) {$\vdots$};
  \node[anchor=east] at (\xo,\yK) {$K$};

  % the spine
  \draw[black!50, fill=black!8, rounded corners=3pt] (-0.8,1.95) rectangle (0.8,\yK-0.18);
  \node[anchor=south] at (0,2.0) {spine, kept by rank 0};

  % subtrees: apex, rank, half-width at the last orbital. Dealt at k+1 to the right, at k+2 to
  % the left, the heaviest nearest the spine; the turn reverses at the next orbital.
  \foreach \xa/\ya/\g/\hw in {1.4/-0.7/0/0.44, 2.4/-0.7/1/0.44, 3.4/-0.7/2/0.44,
      -1.4/-1.4/2/0.38, -2.4/-1.4/1/0.38, -3.4/-1.4/0/0.38} {
    \fill[g\g!38] (\xa,\ya) -- (\xa-\hw,\yK) -- (\xa+\hw,\yK) -- cycle;
    \node[font=\scriptsize] at (\xa,\yK+0.2) {rank \g};
    \fill[g\g] (\xa-\hw,\ybar-0.13) rectangle (\xa+\hw,\ybar);
    \draw[->, semithick] (\xa,\ybar-0.17) -- (\xa,\yres+0.03);
  }
  \node[align=center, font=\scriptsize, black!75] at (2.45,0.42) {dealt at $k{+}1$\\heaviest first: ranks 0, 1, 2};
  \node[align=center, font=\scriptsize, black!75] at (-2.4,-0.3) {dealt at $k{+}2$, in reverse:\\ranks 2, 1, 0};

  % inside the spine: a tree, every node has one parent
  \foreach \x/\o in {-0.6/0, -0.2/{\uparrow}, 0.2/{\downarrow}, 0.6/{\uparrow\downarrow}} {
    \draw[edge] (0,1.6) -- (\x,1.05);
    \node[font=\scriptsize, black!75] at (\x,0.78) {$\o$};
  }
  \node[black!60] at (0,0.4) {$\vdots$};
  \draw[edge] (-\xs,0.22) -- (-\xs,0) (\xs,0.22) -- (\xs,0);
  \draw[edge] (-\xs,0) -- (-\xs,-\dy) (-\xs,0) -- (\xs,-\dy);
  \draw[edge] (-\xs,-\dy) -- (-\xs,-3*\dy) (\xs,-\dy) -- (\xs,-3*\dy);
  \draw[edge] (-\xs,-3*\dy) -- (-\xs,-4*\dy) (-\xs,-3*\dy) -- (\xs,-4*\dy);
  \draw[edge] (-\xs,-4*\dy) -- (-\xs,-4*\dy-0.25) (\xs,-4*\dy) -- (\xs,-4*\dy-0.25);
  \draw[edge] (-\xs,\yK+0.25) -- (-\xs,\yK) (\xs,\yK+0.25) -- (\xs,\yK);
  \node[black!60] at (0,\yK+0.55) {$\vdots$};
  \fill[g0] (-0.5,\ybar-0.13) rectangle (0.5,\ybar);
  \draw[->, semithick] (0,\ybar-0.17) -- (0,\yres+0.03);

  % pruned branches
  \foreach \xa/\ya/\xb/\yb in {\xs/-2.1/0.6/-2.45, -\xs/-2.8/-0.6/-3.15} {
    \draw[black!45, dashed] (\xa,\ya) -- (\xb,\yb);
    \node[black!60, inner sep=0pt] at (\xb,\yb) {$\times$};
  }

  % dealt branches, labeled with the occupation of the next orbital
  \foreach \xa/\ya/\xb/\yb/\o/\p in {\xs/0/1.4/-0.7/{$\downarrow$}/0.62, \xs/0/2.4/-0.7/{$\uparrow$}/0.8,
      \xs/0/3.4/-0.7/{$\uparrow\downarrow$}/0.84, -\xs/-0.7/-1.4/-1.4/{$0$}/0.55,
      -\xs/-0.7/-2.4/-1.4/{$\downarrow$}/0.78, -\xs/-0.7/-3.4/-1.4/{$\uparrow$}/0.84} {
    \draw[deal] (\xa,\ya) -- node[occ, pos=\p] {\o} (\xb,\yb);
  }

  % nodes on top
  \fill[black!60] (0,1.6) circle (0.06);
  \foreach \x in {-0.6,-0.2,0.2,0.6} { \fill[black!60] (\x,1.05) circle (0.06); }
  \foreach \y in {0,-0.7,-1.4,-2.1,-2.8,\yK} {
    \fill[black!60] (-\xs,\y) circle (0.06);
    \fill[black!60] (\xs,\y) circle (0.06);
  }
  \foreach \xa/\ya/\g in {1.4/-0.7/0, 2.4/-0.7/1, 3.4/-0.7/2, -1.4/-1.4/2, -2.4/-1.4/1, -3.4/-1.4/0} {
    \fill[g\g] (\xa,\ya) circle (0.075);
  }

  % merged result
  \node[anchor=east, align=right, font=\scriptsize] at (\xo,\ybar-0.07) {sorted\\lists};
  \fill[black!22] (-3.84,\yres-0.30) rectangle (3.84,\yres);
  \draw[black!65] (-3.84,\yres-0.30) rectangle (3.84,\yres);
  \node at (0,\yres-0.15) {determinants above the threshold, sorted};
  \node[anchor=north] at (0,\yres-0.36) {merged on rank 0};
\end{tikzpicture}}
  \caption{Search for the leading determinants on three ranks. Each level of the tree is one
  orbital, and every partial occupation branches into the four occupations of the next
  orbital, $0$, ${\uparrow}$, ${\downarrow}$ and ${\uparrow\downarrow}$, which label the
  edges. A crossed branch has a norm below the threshold and is dropped together with
  everything below it. Rank 0 keeps the spine, a fixed number of the heaviest partial
  occupations at every orbital. From orbital $k$ on, where the partial occupations
  outnumber the spine, it deals the remaining ones to the ranks one at a time, starting
  with the heaviest, and reverses the order of the ranks at every orbital. The ranks
  complete their subtrees on their own. Each colored subtree stands for the many subtrees
  that a rank receives. Each rank sorts its determinants, and rank 0 merges the sorted
  lists.}
  \label{fig:dets-split}
\end{figure}
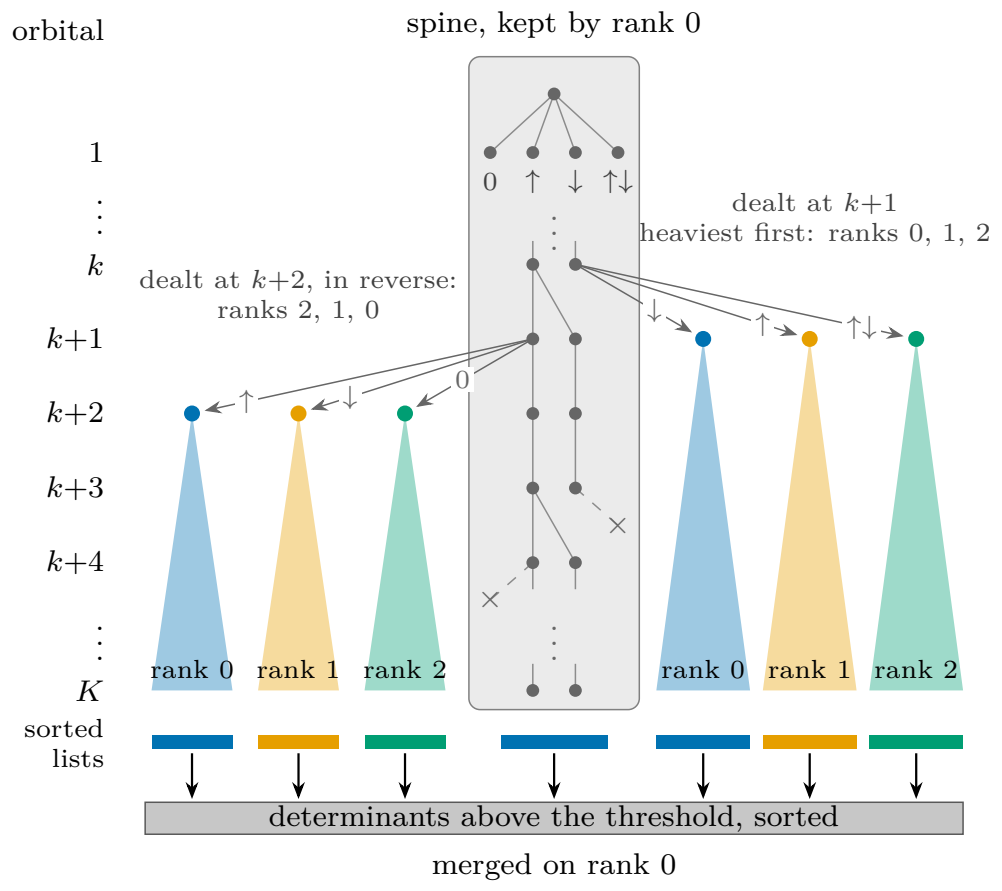

Over several ranks the search tree is divided as in \cref{fig:dets-split}. The squared norm of
a partial occupation is the weight of all determinants that begin with it, and it serves as the
estimate of the work below it. Rank 0 follows a narrow spine through the tree: at every orbital
it keeps a fixed number of the heaviest partial occupations and passes the other surviving ones
to the ranks, the heaviest first and in turn, so that the ranks receive nearly equal weight. A
dominant branch, such as that of the Hartree--Fock determinant, is in this way shared out as it
branches instead of going to one rank whole. Each rank then completes its subtrees without
further communication and sorts its determinants, and rank 0 merges the sorted lists. Every
partial occupation belongs to exactly one rank and is pruned by the same bound as on a single
rank, so the division loses no determinant above the threshold.

For a spin-adapted state the search works on the uncoupled form (\cref{sec:impl-su2}) and
returns the determinants of its member with $S_z = S$. For a singlet this is the state itself.

The determinants serve to characterize a state, to compare it with configuration interaction
vectors, and to hand it to methods that work with determinants, such as selected configuration
interaction or quantum Monte Carlo, as a starting or trial wave function.

\begin{figure}[!htb]
  \centering
  \includegraphics[width=\columnwidth]{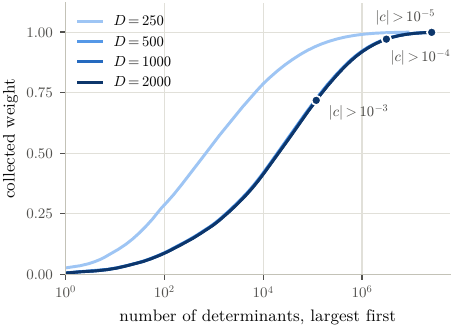}
  \caption{Slater determinants of the [2Fe-2S] ground state: weight collected by the $n$
    determinants with the largest coefficients $c$ (U(1) symmetry). The state is optimized at
    $D=2000$ and then projected to the smaller bond dimensions. The points mark the three sets
    that the search returns for the thresholds $10^{-3}$, $10^{-4}$ and $10^{-5}$ at
    $D=2000$. The curves for $D=500$, 1000 and 2000 lie on top of each other, their collected
    weights differ by less than 0.01. Only the curve for $D=250$ is separate.}
  \label{fig:dets}
\end{figure}

\Cref{fig:dets} shows how the weight of the [2Fe-2S] ground state is spread over the
determinants. The state is far from a single determinant. The two largest coefficients, 0.073
each, belong to the two determinants with five electrons of one spin on one iron atom and five
of the other spin on the second, and each of them holds 0.5\,\% of the weight. The
$1.2\times10^{5}$ determinants with a coefficient above $10^{-3}$ collect 72\,\% of the weight,
the $3.1\times10^{6}$ above $10^{-4}$ collect 97\,\%, and the $2.6\times10^{7}$ above $10^{-5}$
collect 99.9\,\%. The search for the last set takes 2.3\,s on one GPU. From $D=500$ on the
curve no longer changes with the bond dimension. At
$D=250$ the state is the mixture of spin states of \cref{sec:feat-excited}, with
$\langle S^2\rangle = 2.0$, and its weight is concentrated on fewer determinants.

\subsection{Orbital optimization}
\label{sec:feat-casscf}

With the solver of \cref{sec:feat-interface}, \code{} performs DMRG-CASSCF calculations. PySCF
optimizes the orbitals, and \code{} supplies the energy and the RDMs of the active space in
every macroiteration. Restricted orbitals are supported with both symmetries, unrestricted orbitals with U(1) symmetry. State-averaged CASSCF uses the state-averaged RDMs of \cref{sec:feat-excited}, and with SU(2) symmetry it can average states of different total spin, one calculation per spin. Two measures keep the cost of
the many DMRG calculations down. The inner iterations of the orbital optimizer call the solver
with fewer sweeps and a looser threshold. And the state of one macroiteration can seed the
next: after a small rotation of the orbitals the renormalized bases are still a good starting
point, so they are kept, and only the operators are rebuilt from the new integrals.

\begin{figure}[!htb]
  \centering
  \includegraphics[width=\columnwidth]{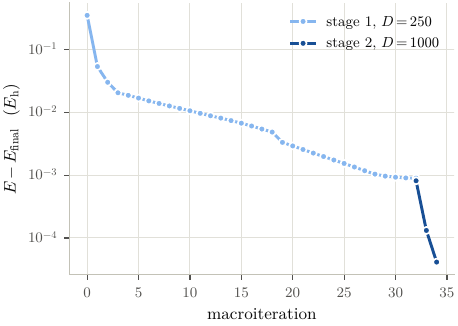}
  \caption{DMRG-CASSCF calculation for the [2Fe-2S] cluster in two stages, with SU(2)
    symmetry: energy after every macroiteration, relative to the final energy. Macroiteration 0
    is the calculation in the published orbitals. The last two macroiterations of stage 2 lie
    within $10^{-9}\,E_{\mathrm h}$ of the final energy and are not shown.}
  \label{fig:casscf}
\end{figure}

The [2Fe-2S] cluster is an example with a large orbital space, namely 662 basis functions in the
basis of the published model. The published orbitals are localized and not optimized for the
active space, which means that the optimization leads to an energy improvement by as much as 
0.35\,$E_{\mathrm h}$. To speed up the optimization we split it in two stages, a common procedure in
DMRG-CASSCF calculations (\cref{fig:casscf}). Stage 1 optimizes the orbitals at $D=250$ and
takes 32 macroiterations, after which the energy is already within $10^{-3}\,E_{\mathrm h}$ of the
final value. Stage 2 starts from these orbitals at $D=1000$ (\cref{lst:casscf}) and converges
in four macroiterations, in which the energy drops by another $8\times10^{-4}\,E_{\mathrm h}$.

A solver that truncates leaves noise in the energy and in the orbital gradient, and the
thresholds of the orbital optimizer have to allow for it. In the example a stage ends when the
energy changes by less than $5\times10^{-5}\,E_{\mathrm h}$ (stage 1) or
$10^{-6}\,E_{\mathrm h}$ (stage 2) from one macroiteration to the next and the norm of the
orbital gradient is below 0.1. The result is converged within these thresholds. At the end of
stage 2 the norm of the orbital gradient is 0.008.

Seeding pays off in every macroiteration. With it, a macroiteration of stage 2 takes 8 sweeps,
against 12 when every DMRG calculation starts from scratch, and the two variants end at the same
energy to $10^{-10}\,E_{\mathrm h}$. On one GPU stage 1 takes 82 minutes, about 2.5 minutes per
macroiteration, and stage 2 takes 19 minutes with seeding and 24 without. The same calculation also
runs unchanged on several GPUs. On four GPUs stage 2 with seeding takes the same four
macroiterations, ends at the same energy to $10^{-10}\,E_{\mathrm h}$ and takes 14 minutes.

\subsection{Convergence aids and restart}
\label{sec:feat-aids}

\paragraph{Orbital order.} The bond dimension required to achieve a given accuracy may depend significantly on the
order of the orbitals in the chain, an observation that has been identified early on in DMRG for quantum chemistry \cite{olivares2015,baiardi2020}. \code{} offers two automated orders. Both sort the orbitals by
the Fiedler vector of a graph whose edge weights measure how strongly two orbitals
interact~\cite{barcza2011,olivares2015}. The first takes the weights from the exchange
integrals and comses with basically no extra costs. The second takes them from the mutual information of a preliminary
calculation at small bond dimension. With SU(2) symmetry the preliminary calculation is spin-adapted as well. A permutation can also be given directly. The program
applies the order internally and returns all results in the orbital order of the input.

\begin{figure}[!htb]
  \centering
  \includegraphics[width=\columnwidth]{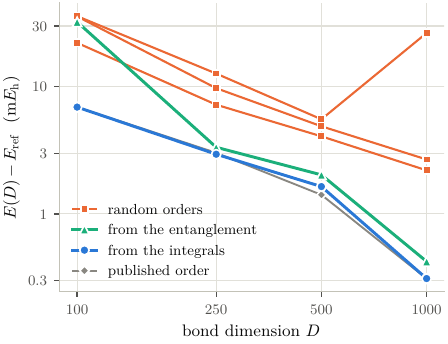}
  \caption{Energy error of the [2Fe-2S] ground state against the bond dimension for different
    orders of the orbitals (U(1) symmetry): the order of the published integrals, random
    relabelings of the orbitals, and the two automatic orders, both computed from the first of
    the random ones. The reference is the energy published with the integrals.}
  \label{fig:order}
\end{figure}

\Cref{fig:order} shows the effect for the [2Fe-2S] cluster. The published integrals come in an
order that suits DMRG. When the orbitals are relabeled at random, the error of the energy at a
given bond dimension is two to nine times larger. One of the three calculations at $D=1000$
ends close to the high-spin state of \cref{fig:ladder}, with $\langle S^2\rangle = 27$ and
26\,m$E_{\mathrm h}$ above the singlet. From the random labels, the order from the exchange
integrals restores the published order exactly, in reverse, and with it the energies. The order
from the mutual information depends on its preliminary calculation, which runs at an eighth of
the bond dimension. At $D=100$ this is too small and the order is no better than a random one.
From $D=250$ on it comes close to the order from the integrals.

\paragraph{Starting guess and noise.} A two-site sweep can settle in a state that is not the
ground state and that further sweeps do not leave. \code{} offers two means that influence where
a calculation settles. The first chooses the starting point: the leading determinants of a
configuration interaction calculation are converted into a matrix product state that seeds the
first sweep~\cite{li2021}. The second perturbs the optimization: during the first sweeps a term built from the
operators of the block is added to the density matrix of the decimation~\cite{white2005}, with
an amplitude that the user sets per sweep and that is zero in the final sweeps. Both are available only with U(1) symmetry so far. A spin-adapted calculation can instead start
from the saved bases of an earlier spin-adapted calculation. The calculations
with U(1) symmetry of this section use this term, with an amplitude of $10^{-4}$ in the first
four sweeps and $10^{-5}$ in the next two.

\paragraph{Checkpoints and export.} \code{} can save the renormalized bases after every sweep.
A calculation that was stopped, for example at the time limit of a job, continues from the last
saved sweep and gives the same result as the uninterrupted calculation.  The saved bases can also start a new calculation of the same symmetry. The matrix product state can be written to a file, read back as the starting point of another
calculation, and converted to and from a matrix product state of the quimb
library~\cite{gray2018}. Its tensors are then the block-sparse fermionic arrays of
\texttt{symmray}~\cite{gao2025fermionic, symmray}. Overlaps between two states are computed by the engine. A spin-adapted state is exported in its uncoupled form (\cref{sec:impl-su2}), since the symmetric arrays of \texttt{symmray} are Abelian. The exported state cannot start a spin-adapted calculation, which restarts from its saved bases instead.
\section{Conclusions}
\label{sec:conclusion}

We have presented \code, an open-source program for ab initio DMRG designed for machines with
many GPUs.
Its design follows from one observation: device memory is the scarce resource of a large
calculation. The pair operators stay on the devices, and each process holds only its own share
of them. The complementary operators are formed at the bond that needs them and are never
stored. A memory plan on which the processes agree chooses at every bond what the devices, the
hosts and the disks hold. Together with matrix--vector products and a renormalization that run
as batched matrix multiplications on the devices, this makes bond dimensions well above $10^4$
and active spaces of 100 orbitals accessible on today's GPU machines.

The release version scales well into the hundreds of GPUs and, on CPU nodes, to thousands of
cores. The matrix--vector products, by far the largest part of a sweep, scale almost ideally.
Communication between the processes takes at most 3\,\% of a sweep, and the GPUs become more
efficient as the bond dimension grows, so the largest calculations are the ones that use the
machine best. The largest calculation, the P-cluster at $D = 28\,000$ on 512 GPUs, sustains
9.6\,PFLOPS of useful operations at twice the bond dimension reached previously with
\textsc{Focus}~\cite{xiang2024}. What limits the scaling beyond this are a few small parts of a
sweep whose cost per process stays the same when processes are added. In the one-dimensional
distribution they are fixed by its layout, because every process forms and receives a share of
every complementary operator. The two-dimensional distribution already scales nearly as well up
to 512 GPUs in its first version, balances the work over the rows of its grid and divides the
completion of $P$ and $Q$ by the number of its columns. Running it with more columns and with
the completion of $S$ by columns, and extending it to spin-adapted calculations, is our route
from hundreds to thousands of GPUs.

The application to the reaction chemistry of cytochrome P450 with the drug molecule warfarin shows the program perform at scale in production calculations. Reactant and transition
state of the hydroxylation of warfarin were computed in an active space of 50 orbitals at a
bond dimension of 7000, with unrestricted orbitals and with energies extrapolated to zero
discarded weight. Active spaces of up to 100 orbitals at bond dimensions up to 12\,000 will
follow. Such calculations make large active spaces available for the study of reaction
mechanisms, where several structures have to be treated within the same computational framework.

\code{} is more than a ground-state solver. Excited states, reduced density matrices, orbital
entanglement, the leading Slater determinants of a state, orbital optimization with PySCF,
automatic orbital orders and restart are part of the release. They are driven from Python in a
few lines, and the sweeps and measurements behind them run on the GPUs and over several
processes. The program is under continuous development, and further capabilities are in
progress. To name a few, these are the optimization of single excited states and transition
density matrices, three-particle reduced density matrices as the basis of methods for the
correlation outside the active space, orbital entanglement for every state of a state-averaged
calculation, and the starting guess from configuration interaction and the perturbative noise
for spin-adapted calculations. Our aim is a
universal DMRG program for quantum chemistry, one that covers the path from the integrals to
the analyzed wave function on CPUs and on GPUs alike. The code is open source, and we invite
the community to use and contribute to it.

\section*{Code availability}
\code{} is open source under the Apache-2.0 license and available at
\url{https://github.com/Algorithmiq/exa-dmrg}. The repository includes documentation that
introduces DMRG and describes the implementation, the scripts of all examples shown in this
paper, and scaling plots that are kept up to date with the development of the code.

\section*{Author contributions}
F.L. and F.T. developed \code{}, and F.L. led the development. R.D.R.E. prepared the public
release of the code and its distribution as Python packages. S.K., F.P. and R.D.R.E. advised on
technical and theoretical questions during the development. P.A.B.H. provided the
active spaces and Hamiltonians of the cytochrome P450 study in \cref{sec:p450}. F.L. wrote the
manuscript. All authors reviewed and revised the manuscript.

\section*{Acknowledgments}
We thank Esperanza Cuenca G\'omez (NVIDIA) for helpful discussions during our work on the scaling of
the code and on its optimization for NVIDIA GPUs. 
The authors are grateful to Janko \v{C}ivi\'c (KU Leuven)\ who optimized the structures of the warfarin metabolism reaction as part of his internship at Algorithmiq. 
We acknowledge the EuroHPC Joint Undertaking for awarding the project EHPC-EXT-2026E01-156
access to the EuroHPC supercomputers JUPITER, hosted by the J\"ulich Supercomputing Centre in
Germany, and MareNostrum~5, hosted by the Barcelona Supercomputing Center in Spain, through a
EuroHPC Extreme Scale Access call.

\section*{AI usage statement}
Several Claude models (Anthropic) were used to optimize the
implementation of the code, and
Claude Fable 5.2 (Anthropic) to assist with the writing of this manuscript. All code was
reviewed, tested and validated by the authors, and all text was reviewed and verified by them.
The authors take full responsibility for the code and for the content of this paper.

\bibliographystyle{unsrtnat}
\bibliography{references}

\clearpage
\onecolumn
\setcounter{section}{0}\setcounter{figure}{0}\setcounter{table}{0}\setcounter{equation}{0}
\renewcommand{\thesection}{S\arabic{section}}
\renewcommand{\thefigure}{S\arabic{figure}}
\renewcommand{\thetable}{S\arabic{table}}
\renewcommand{\theequation}{S\arabic{equation}}
\renewcommand{\theHsection}{S\arabic{section}}
\renewcommand{\theHfigure}{S\arabic{figure}}
\renewcommand{\theHtable}{S\arabic{table}}
\renewcommand{\theHequation}{S\arabic{equation}}
\begin{center}
  {\LARGE Supporting information}\\[6pt]
  {\large \code: Toward exascale quantum chemistry with open-source DMRG, from a single CPU
  core to hundreds of GPUs}
\end{center}
\vspace{1em}
\section{Further scaling series}
\label{app:scaling}

The series of \cref{sec:bench} that are not in \cref{fig:bench-main}, one figure per series. Left:
time per sweep with the efficiency at every point, relative to the most efficient point, and the
ideal scaling through that point (dashed). Right: the sweep by phase.

\begin{figure}[!htb]
  \centering
  \includegraphics[width=\textwidth]{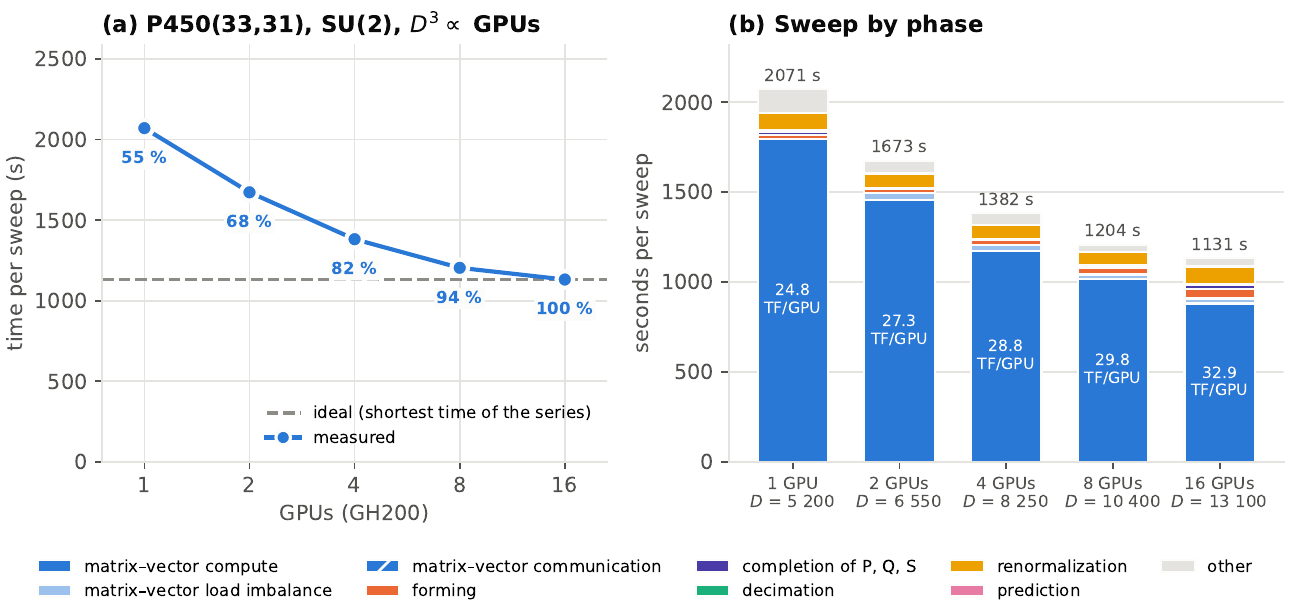}
  \caption{Weak scaling of P450(33,31) with SU(2) on GPUs, $D = 5200$ to 13\,100
    on 1 to 16 GPUs ($D^3 \propto n$), \cref{sec:bench-weak}. (b) in seconds, with the GEMM rate
    per GPU of the matrix--vector products in the bar.}
  \label{fig:app-cpdie}
\end{figure}

\begin{figure}[!htb]
  \centering
  \includegraphics[width=\textwidth]{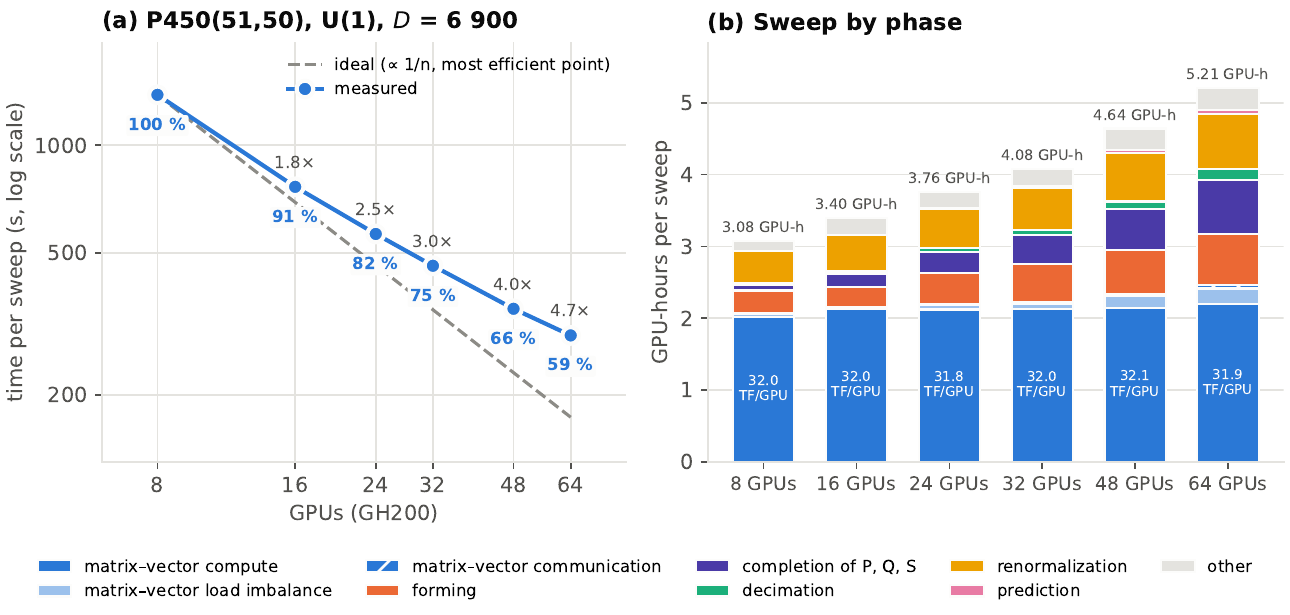}
  \caption{Strong scaling of P450(51,50) with U(1) on GPUs, $D = 6900$ on 8 to 64 GPUs,
    \cref{sec:bench-strong}. (a) also the speed-up over 8 GPUs. (b) in GPU-hours, the number of
    GPUs times the time, so that ideal strong scaling keeps the bars level.}
  \label{fig:app-p450}
\end{figure}

\begin{figure}[!htb]
  \centering
  \includegraphics[width=\textwidth]{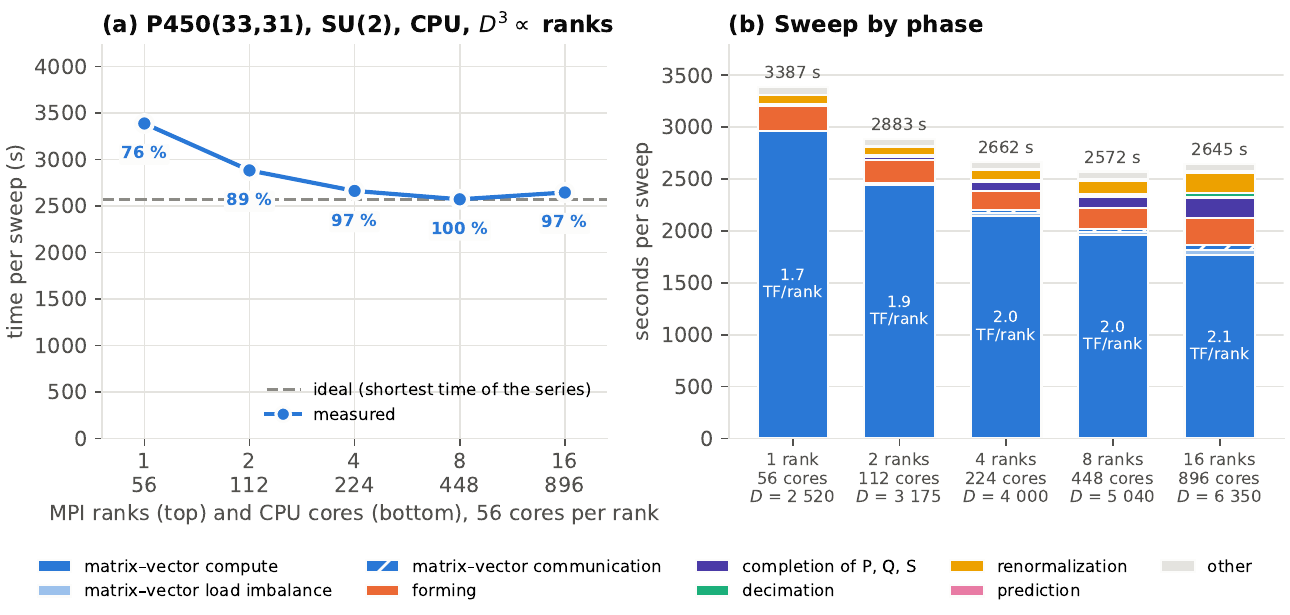}
  \caption{Weak scaling of P450(33,31) with SU(2) on the CPU nodes of MareNostrum~5,
    $D = 2520$ to 6350 on 1 to 16 ranks of 56 cores each ($D^3 \propto n$), \cref{sec:bench-cpu}.
    (b) in seconds, with the rate per rank of the matrix--vector products in the bar, their GEMM
    work over their computation time. On one rank the work is not recorded and is extrapolated
    from the points on 2 and 4 ranks.}
  \label{fig:app-cpu-weak}
\end{figure}

\begin{figure}[!htb]
  \centering
  \includegraphics[width=\textwidth]{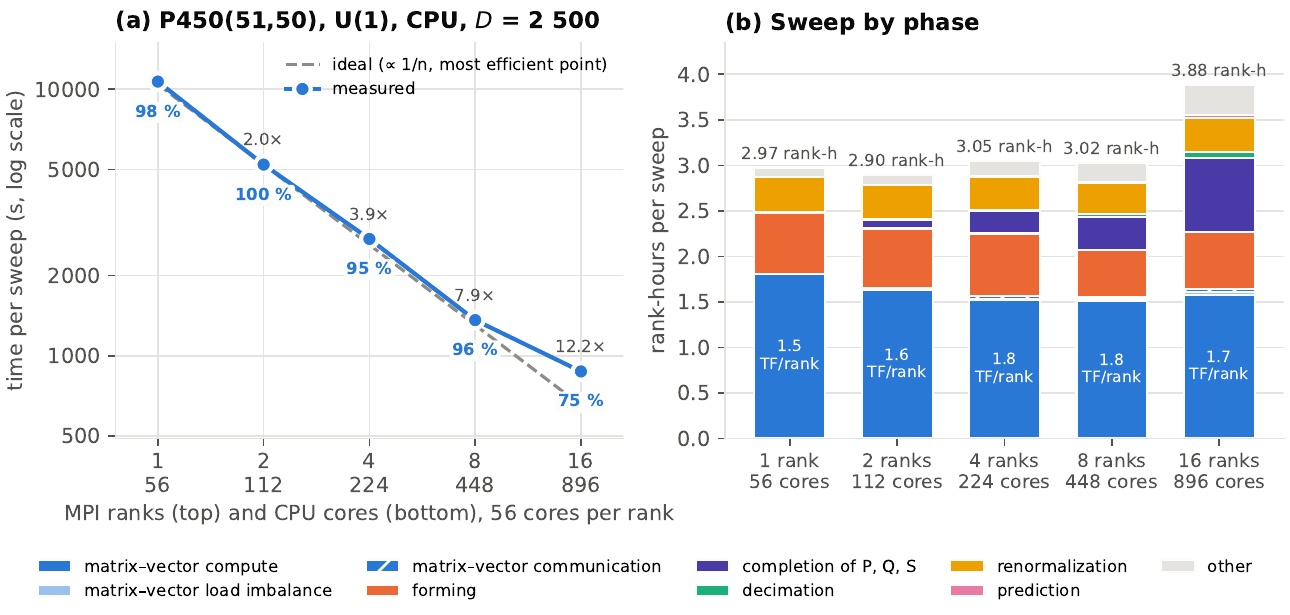}
  \caption{Strong scaling of P450(51,50) with U(1) on the CPU nodes of MareNostrum~5,
    $D = 2500$ on 1 to 16 ranks of 56 cores each, \cref{sec:bench-cpu}. (b) in rank-hours, the
    number of ranks times the time, with the rate per rank as in \cref{fig:app-cpu-weak}. On one
    rank the work is that of the point on 2 ranks.}
  \label{fig:app-cpu-strong}
\end{figure}

\end{document}